\documentclass[10pt,a4paper]{article}
\usepackage[utf8]{inputenc}
\usepackage[english]{babel}
\usepackage[toc]{appendix}
\usepackage{graphpap,amsmath,amsfonts,amssymb,epsfig,makeidx,graphicx,textcomp,cite,citeall,authblk,subfloat,float,xcolor} 
\usepackage[left=1.5cm, right=1.5cm, top=1.5cm, bottom=1.5cm]{geometry}

\author[1]{{Soham Chandra} \thanks{E-mail addresses: soham.rs@presiuniv.ac.in ; sohamc07@gmail.com}}

\affil[1]{\textit{\normalsize{Department of Physics, Presidency University, 86/1 College Street, Kolkata -700 073, India}}}

\vspace{20pt}

\title{\textbf{Effects of site dilution on Compensation in Ising Spin-1/2 trilayered triangular Ferrimagnets with non-equivalent planes}}

\date{} 
\begin{document}
	\maketitle	
	\begin{abstract}
		Using Monte Carlo simulations with the Metropolis algorithm, the magnetic and thermodynamic behaviours of a spin-1/2, trilayered ferrimagnetic system on \textit{triangular monolayers} with quenched nonmagnetic impurities are studied. Two different theoretical atoms, A and B, make up the \textit{ABA} and \textit{AAB} types of distinct configurations. Like atoms (A-A and B-B) interact ferromagnetically, while unlike atoms (A-B) interact antiferromagnetically. Only the A-layers are randomly site-diluted with dilution percentages ranging from 5\% to 45\%. Such diluted magnetic thin systems exhibit \textit{magnetic compensation} which depends sensitively on the concentration of impurities. The phase diagram in the Hamiltonian parameter space related to the occurrence of magnetic compensation phenomenon and the effect of site dilution is discussed in detail. Special attention is given to the mathematical dependencies of compensation temperature on the concentration of nonmagnetic impurities. Depending upon the concentration of nonmagnetic impurities, the compensation and critical points shift with the equilibrium magnetic behaviours changing between distinct ferrimagnetic behaviours. For each combination of the coupling strengths, with values of the impurity concentration above a threshold, compensation appears where previously was absent. Suggested mathematical formulae show how threshold impurity concentration relies on Hamiltonian parameters. 
	\end{abstract}


\newpage
\twocolumn
\section{Introduction}
\label{sec_intro}
After the discovery of ferrimagnetism in 1948 \cite{Neel,Cullity}, a ferrimagnet is often modelled as a combination of two or more magnetic substructures, e.g., sublattices, sublayers or subsets of atoms. Among them, a certain class of layered ferrimagnets have generated considerable attention in recent times as they show \textit{compensation effect}. Such thin magnetic systems, for which one dimension is greatly reduced than the other two, serve as a bridge between monolayer (2D) and bulk (3D) versions of a magnetic material. Different magnetic responses to the variation of temperature of component sublayers, when combined, lead to the appearance of compensation points, i.e., temperatures lower than the critical point for which magnetization of the total system is zero but the sublayers retain magnetic ordering (see references \cite{Diaz1,Diaz2,Chandra1,Chandra2,Chandra3}). Such layered ferrimagnets with ferromagnetic and antiferromagnetic interactions have revealed interesting magnetic responses and phase diagrams. The layered ferrimagnets with interlayer antiferromagnetic coupling between adjacent ferromagnetic layers have been used in giant magnetoresistance (GMR) \cite{Camley}, magneto-optical recordings \cite{Connell}, the magnetocaloric effect \cite{Phan} and spintronics \cite{Grunberg}. That is why both theoretical and experimental investigations on such ferrimagnets are important. The experimental realization of bilayer \cite{Stier}, trilayer \cite{Smits,Leiner}, and multilayer \cite{Chern,Sankowski,Chung,Samburskaya,Pradhan,Maitra} systems, with desired characteristics, has become a reality as thin film growth techniques, e.g. molecular-beam epitaxy (MBE) \cite{Herman}, metalorganic chemical vapour deposition (MOCVD) \cite{Stringfellow}, pulsed laser deposition (PLD) \cite{SinghRK}, and atomic layer deposition (ALD) \cite{Leskela,George} have been developed and extensively used with time.\\

Extensive search for novel, magnetically compensated materials exhibiting better performance is centered around the manipulation and control of compensation points \cite{Sandeman,Manosa}. As a result, theoretical magnetic models are useful for providing physical insights into the compensation phenomenon and describing experimental data. Now let us review a few of the popular theoretical techniques for treating spin models. References \cite{Diaz1,Amaral,Franco,Dong,Oliveira1,Amaral2,Franco2,Pelka} demonstrate the versatility of the mean-field analysis as a very successful semi-analytical method for establishing the thermodynamics of the magnetic systems. Furthermore, there are only a few analytical techniques that are exactly solvable for different spin-systems, such as the generalized classical transfer matrix method and decoration-iteration mapping \cite{Canova,Pereira,Ohanyan,Strecka,Galisova}, Bethe ansatz-based quantum transfer matrix and nonlinear integral equations method \cite{Ribeiro,Trippe}, the Jordan-Wigner transformation \cite{Zhitomirsky,Topilko} etc. In \cite{Chang}, the dynamic magnetic properties of antiferromagnetic/ferromagnetic $YMnO_{3}$/FM bilayer under a time-dependent magnetic field are studied by Monte Carlo simulation on a mixed-spin $(5/2, 2, 3/2)$ Ising model. By Monte Carlo Metropolis algorithm, in \cite{Guru}, \textit{the effect of next nearest neighbour interactions} on compensation temperature and phase transition is investigated in a trilayered ferrimagnetic system on a square lattice. This technological aspect e.g. the magnetocaloric effect in the spin-$1$ Blume-Capel model is theoretically studied in \cite{Oliveira2}, using the mean-field theory from the Bogoliubov inequality. \\


Let us now look at some recent developments in diluted magnetic systems. The effect of site dilution on compensation and critical temperatures of a two-dimensional mixed spin-1/2 and spin-1 system has been studied using Monte Carlo simulations, in \cite{Aydiner}. In \cite{Yuksel}, the effects of dilution in a cylindrical magnetic nanowire system composed of ferromagnetic core and shell layers were investigated using effective field theory (EFT) with correlations, with both ferromagnetic and antiferromagnetic exchange couplings at the core-shell interface and in the antiferromagnetic nanowire case, a compensation point could be introduced by bond dilution at the surface. In \cite{Diaz3}, Monte Carlo simulation with Wolff cluster algorithm is employed to investigate the thermodynamic and magnetic properties of a \textit{site diluted} spin-1/2 Ising multilayered ferrimagnet. The investigated system is made up of non-equivalent planes with quenched site diluted alternating planes having the dominant in-plane coupling. In \cite{Vatansever}, the effect of nonmagnetic impurities and roughness on the finite temperature magnetic properties of core-shell spherical nanoparticles with anti-ferromagnetic interface coupling is investigated. In \cite{Diaz4}, a Monte Carlo study of the magnetic properties of an Ising multilayer ferrimagnet is performed with two kinds of non-equivalent planes, one of which is site-diluted. In \cite{Fadil}, the magnetic properties of a diluted trilayered graphene structure with non-equivalent planes with alternating spins $1$ and $3/2$ are studied. In \cite{Qajjour}, the effect of site dilution on the magnetic properties of a mixed spin honeycomb nano-lattice (with two sublattices, one with spin-3/2 and the other with spin-5/2) is investigated using Monte Carlo simulations with the Metropolis algorithm.\\

So previous results in the literature indicate that the magnetic properties of the thin magnetic systems with nonmagnetic impurities are interesting compared to their pristine counterparts. Such studies provide better insight into the physics behind real magnetic quasi-3D systems. After experimentally realizing artificial gauge potentials, it is now possible to engineer relevant spin interactions in quantum simulators \cite{Buluta} using ultracold bosonic quantum gases \cite{Lewenstein} in optical lattices on a triangular lattice geometry \cite{Struck}. This methodology is successfully implemented by a Penning trap apparatus with laser-cooled ${}^{9}Be^{+}$ ions ($\sim 300$ spins), naturally forming a stable 2D triangular Coulomb crystal \cite{Britton}. Here, each ion is a spin-$1/2$ qubit and in \cite{Biercuk}, the authors used high-fidelity quantum control and a spin-dependent optical dipole force (ODF) to engineer a continuously tunable Ising-type spin-spin coupling. Motivated by these advances, an equilibrium Monte Carlo study on compensation has been performed on a spin-$1/2$ trilayered triangular system \cite{Chandra3}. But how the quenched disorder e.g. site-dilution influences the compensation and critical properties of the trilayered triangular Ising system is still unresolved. In this work, we would use a Monte Carlo approach to investigate a diluted trilayered Ising spin-1/2 ferrimagnet on a triangular lattice similar to the one in \cite{Chandra3}. We, in this article, would mainly focus to find answers to the following:\\
(a) How does the concentration of nonmagnetic impurities at the diluted planes affect the critical and compensation behaviour? \\
(b) How would the phase diagram change with the change in the concentration of impurity?\\
(c) Can we define any mathematical relationship between the physical properties of the system and the concentration of impurities?\\

The rest of the paper is organized as follows: the magnetic model is explained in Section \ref{sec_model}. The strategy of simulation for the trilayered system is presented in Section \ref{sec_simulation}. The numerical results are presented and discussed in Section \ref{sec_results}. The work is summarised in Section \ref{sec_summary}.

\section{Theoretical Model}
\label{sec_model}
\begin{figure}[!htb]
	
	\resizebox{9cm}{!}{\includegraphics[angle=0]{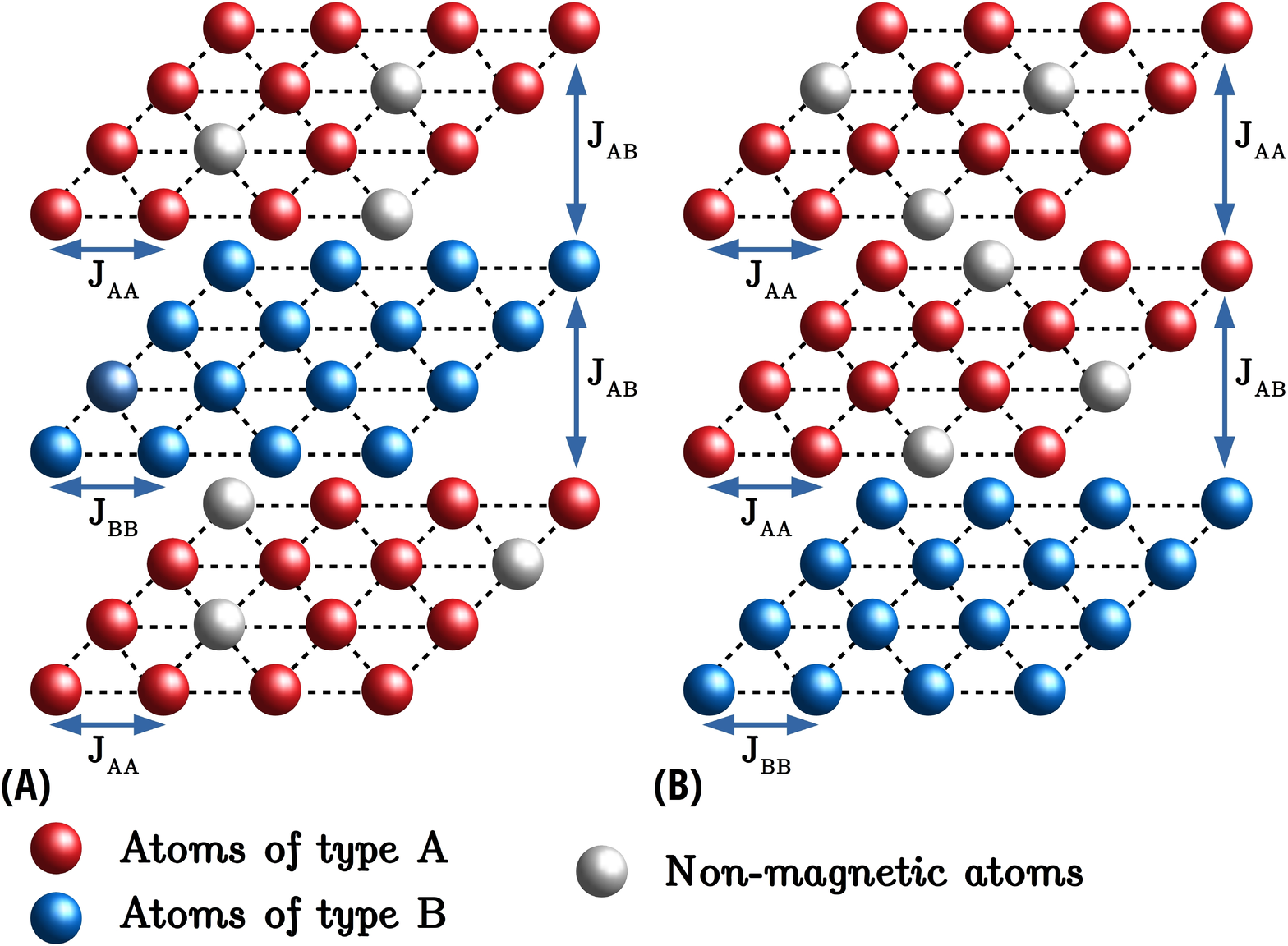}}
	
	\caption{ (Colour Online) Miniaturised versions ($3\times4\times4$) of \textit{site-diluted}: (A) ABA-type and (B) AAB-type triangular trilayered Ising superlattices. Each of the sublattices of the ferrimagnetic systems are formed on triangular lattice. The actual simulation is carried out on a system with $N_{sites}=3\times100\times100$ .} 
	
	\label{fig_1_lattice_structure}
\end{figure}

Our interest in this article is a ferrimagnetic trilayer, in which the spins are located at the sites of the triangular crystalline lattice. The magnetic system is composed of inequivalent
parallel monolayers. In the ABA-type system, alternately stacked planes are populated by either the A or B type of theoretical atoms. Similarly in the AAB-type system, the top and mid-layer are populated by the A-atoms and the bottom layer is populated by the B type of atoms. In both configurations, plane B, with the dominant in-plane coupling is unaffected but the A-planes are site-diluted with the concentration of nonmagnetic atoms being $\rho$. All magnetic atoms are assumed to have spin-$1/2$. The schematic view of the trilayer is presented in Figure \ref{fig_1_lattice_structure}.

The magnetic interactions are limited only between the nearest neighbours and are Ising-like. The nature of magnetic interactions are:(a) A-A $\to$ Ferromagnetic; (b) B-B $\to$ Ferromagnetic; (c) A-B $\to$ Anti-ferromagnetic. The Hamiltonian for such a trilayered ferrimagnetic system, using nearest neighbour Ising mechanics \cite{Ising}, is (with all the $S^{z}$’s are $z$ components of spin moments on lattice sites):
\begin{eqnarray}
\nonumber
H_{ABA} &=& - J_{AA} [ \sum_{<t,t^{\prime}>} (\xi_{t}S_{t}^{z})(\xi_{t^{\prime}} S_{t^{\prime}}^{z}) \\ 
\nonumber
&+& \sum_{<b,b^{\prime}>} (\xi_{b}S_{b}^{z})(\xi_{b^{\prime}}S_{b^{\prime}}^{z})] - J_{BB} \sum_{<m,m^{\prime}>} S_{m}^{z}S_{m^{\prime}}^{z} \\
&-& J_{AB}[ \sum_{<t,m>} (\xi_{t}S_{t}^{z}) S_{m}^{z} +  \sum_{<m,b>} S_{m}^{z} (\xi_{b}S_{b}^{z})]
\label{eq_Hamiltonian_ABA} 
\end{eqnarray}

\begin{eqnarray}
\nonumber
H_{AAB} &=& - J_{AA} [ \sum_{<t,t^{\prime}>} (\xi_{t}S_{t}^{z})(\xi_{t^{\prime}} S_{t^{\prime}}^{z}) + \sum_{<m,m^{\prime}>} (\xi_{m}S_{m}^{z})(\xi_{m^{\prime}}S_{m^{\prime}}^{z})]  \\ 
\nonumber
&-& J_{BB} \sum_{<b,b^{\prime}>} S_{b}^{z}S_{b^{\prime}}^{z}  - J_{AA} \sum_{<t,m>} (\xi_{t}S_{t}^{z}) (\xi_{m}S_{m}^{z}) \\
&-& J_{AB} \sum_{<m,b>} (\xi_{m}S_{m}^{z}) S_{b}^{z}
\label{eq_Hamiltonian_AAB} 
\end{eqnarray}

where $J_{AA}$ is the in-plane coupling strength between nearest neighbours on the A-layers. On similar lines we have, for the B-layer, $J_{BB}$. $J_{AB}$ is the inter-layer nearest
neighbour coupling strengths between the A and B layers. According to the nature of the magnetic interactions: $J_{AA} > 0$ , $J_{BB} > 0$ and $J_{AB} < 0$. $<t,t^{\prime}>$, $<m,m^{\prime}>$ and $<b,b^{\prime}>$ are nearest neighbour summation indices for the top, mid and bottom layers respectively. Similarly $<t,m>$ and $<m,b>$ stand for summations over nearest-neighbor pairs in vertically adjacent layers. We would take $J_{BB}$ to be the dominant coupling strength and use the ferromagnetic coupling ratio $J_{AA}/J_{BB}$ and the antiferromagnetic coupling ratio $J_{AB}/J_{BB}$ to be the controlling parameters. The configurational averages of the occupation probability of A-atoms, for the ABA variant is: $\bar{\xi_{t}}=\bar{\xi_{b}}=(1-\rho)$ and for the AAB variant is: $\bar{\xi_{t}}=\bar{\xi_{m}}=(1-\rho)$, for a given concentration of nonmagnetic atoms, $\rho$. There is no out-of-plane interaction between the top and bottom layers. We will use periodic boundary conditions in-plane and open boundary conditions along the vertical.

\section{Simulation Protocol}
\label{sec_simulation}
To study the magneto-thermal behaviours of the two distinct site-diluted Ising trilayered configurations of Figure \ref{fig_1_lattice_structure}, we have used Monte Carlo simulations with Metropolis single-spin flip algorithm \cite{Landau-Binder, Binder-Heermann}. Each of the three planes has a linear dimension of $100$ sites. So the total sites are $N_{sites}=3\times100\times100$. From the discussions in Section \ref{appendix_comp}, we find the size of the lattice is statistically reliable for our purpose. The $z$ components of spin projections at the site, $i$, is $S_{i}^{z}(S_{i}^{z} = \pm 1/2)$, only participate in the Ising interactions. The system was initiated at a high-temperature paramagnetic phase, with a fraction $\rho$ of randomly selected sites on the A-layers being site-diluted. From the rest, half of the spin projections are in $S_{i}^{z}= +1/2$ and the other half is in $S_{i}^{z}= -1/2$. The B-layer is not diluted i.e. it has an equal population of $S_{i}^{z}= +1/2$ and $S_{i}^{z}= -1/2$. Now, a trial configuration is constructed by reorienting a randomly chosen single spin. This trial configuration is accepted with probability, $p= min \{ 1, e^{-\beta \Delta E} \}$; where $\Delta E$ is the energy difference between the trial and the current configuration. Similar $3L^{2}$ random single-spin updates constitute one Monte Carlo sweep (MCS) of the entire system and this one MCS is the unit of time in this study.

At every temperature step, the system goes through $10^{5}$ MCS. At any temperature step, the last configuration of the system at the just previous temperature acts as the starting configuration. For equilibration, we shall discard the first $2\times10^{4}$ MCS. From the next $8\times10^{4}$ MCS, we would collect data after every $800$ MCS (to minimise the auto-correlation effect) for $100$ \textit{uncorrelated} configurations for relevant physical quantities. The temperatures of the systems are measured in the units of $J_{BB}/k_{B}$ . So the temperatures reported in this article are \textit{effectively dimensionless}. From now onwards, temperature and dimensionless temperature would be used and understood interchangeably. Periodic boundary conditions are used in-plane (i.e. along the x and y axes) and open boundary conditions are used along the vertical (i.e. z axis). The in-plane coupling strength of the B-layer, $J_{BB}$ is considered to be the most dominant one and is set to unity (and also the unit of energy scale). Other two coupling strengths, namely, $J_{AA}$ and $J_{AB}$ are measured with respect to $J_{BB}$ i.e. seven distinct values of $J_{AA}/J_{BB}$ and $\left| J_{AB}/J_{BB} \right| $ are considered, from $0.04$ to $1.0$ with an interval of $0.16$. For each distinct combination of coupling strengths, the time/ensemble averages of the following quantities are calculated at each of the temperature steps $(T)$, in the following manner for both the configurations $(ABA, AAB)$ identically:\\

\textbf{(1) Sublattice magnetizations} $M_{qi}(T)$ for the individual layers, are identically calculated after equilibration, at the $i$-th MCS, by:
\begin{equation}
M_{qi}(T) = \dfrac{1}{L^{2}} \sum_{x,y=1}^{L} \left(S_{qi}^{z}(T)\right)_{xy} 
\end{equation}

The sum extends over all sites in each monolayer with $x$ and $y$ being the co-ordinates of a spin on the $q$-th sublayer and runs from $1$ to $L$
($=100$, in this study). Next, the time (equivalently, ensemble) average (in angular braces) is obtained from the $N$ uncorrelated configurations by:

\begin{equation}
\langle M_{q}(T)\rangle = \dfrac{1}{N} \sum_{i=1}^{N} M_{qi}(T)
\end{equation}
In the above equations, $q$ is to be replaced by $t$, $m$ or, $b$ for top, mid or bottom layers respectively. \\

\textbf{(2) Time averaged total magnetisation},  $M_{tot}(T)$, of the trilayer serves as the order parameter and at temperature, $T$, it is defined as:
\begin{equation}
	M_{tot}(T) = \dfrac{1}{3} \left( \langle M_{t}(T)\rangle + \langle M_{m}(T)\rangle + \langle M_{b}(T)\rangle  \right)
\end{equation}

\textbf{(3) Fluctuation of the order parameter}, $\Delta M (T)$, at temperature $T$ is obtained after equilibration as:
\begin{equation}
	\Delta M(T) = \sqrt{\dfrac{1}{N} \sum_{i=1}^{N} \left[ M_{tot,i}(T) - M_{tot}(T) \right]^{2}}
\end{equation}
where $M_{tot,i}(T)$ is the total magnetisation of the system after the $i$-th uncorrelated MCS.\\ 

\textbf{(4) Canonical average of associative energy per site}, $ E (T)$, at temperature $T$ is obtained after equilibration as:
\begin{equation}
E(T)=\dfrac{ \langle H(T)\rangle }{3L^{2}}
\end{equation}
from the $N$ uncorrelated MCS. Equations \ref{eq_Hamiltonian_ABA} and \ref{eq_Hamiltonian_AAB} are used here.\\

\textbf{(5) The fluctuation of the associative energy per site}, $\Delta E (T)$, at temperature $T$ is obtained after equilibration as:
\begin{equation}
\Delta E(T) = \sqrt{\dfrac{1}{N} \sum_{i=1}^{N} \left[ E_{i}(T) - E(T) \right]^{2}}
\end{equation}
where $E_{i}(T)$ is the associative energy per lattice site of the system after the $i$-th uncorrelated MCS.\\

\textbf{(6) The Binder cumulant}, $U_{L}(T)$ for a given $L$ is defined as \cite{Binder-Heermann}: 
\begin{equation}
U_{L}(T)=1-\dfrac{\langle M_{tot}^{4}(T)\rangle_{L}}{3 \langle M_{tot}^{2}(T)\rangle_{L}^{2}}
\end{equation} 
at dimensionless temperature $T$, after equilibration. We shall use the position of minima of $\dfrac{dU_{L}}{dT}$ about the pseudo-critical point (obtained from the plot of $\Delta E $ versus $T$) to arrive at the value of critical temperature \cite{Ferrenberg}. Only a single system size e.g. $L=100$ in this study is used for this purpose.

At each temperature step, averaging is performed over $10$ different sample realizations (same macroscopic but different microscopic arrangements) for every mentioned quantity.  The estimate of error is obtained via Jackknife method \cite{Newman} and are often smaller than the point markers away from the critical point.
 
\section{Results}
\label{sec_results}
\subsection{Thermodynamic response}
\label{subsec_response}
\begin{figure*}[!htb]
	\begin{center}
		\begin{tabular}{c}
			
			\resizebox{8.75cm}{!}{\includegraphics[angle=0]{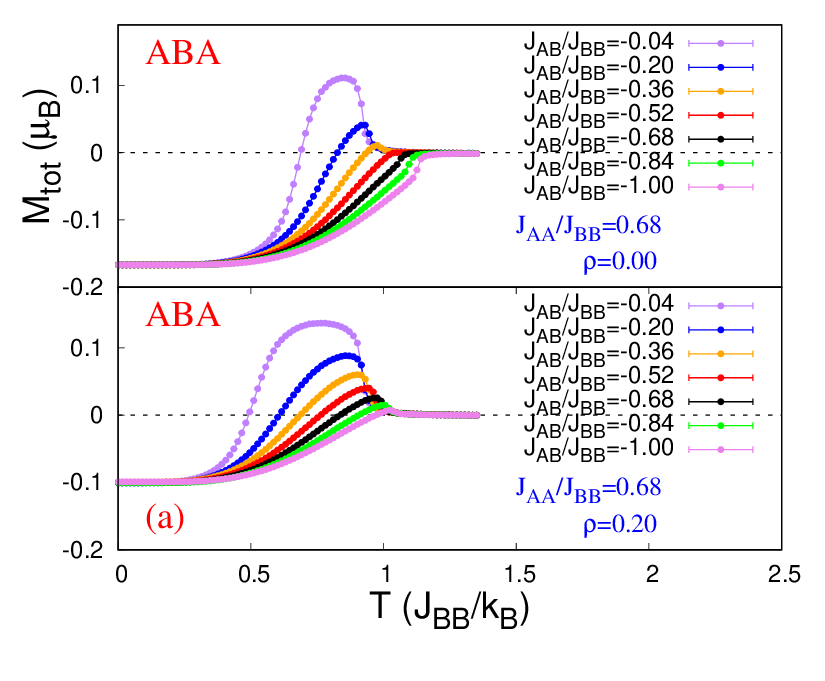}}
			\resizebox{8.75cm}{!}{\includegraphics[angle=0]{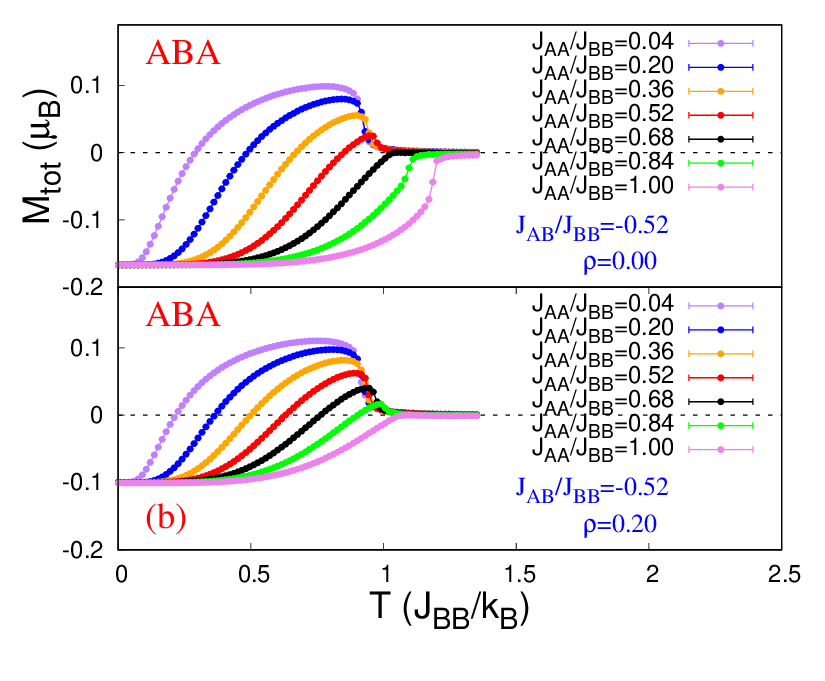}}\\
			
			\resizebox{8.75cm}{!}{\includegraphics[angle=0]{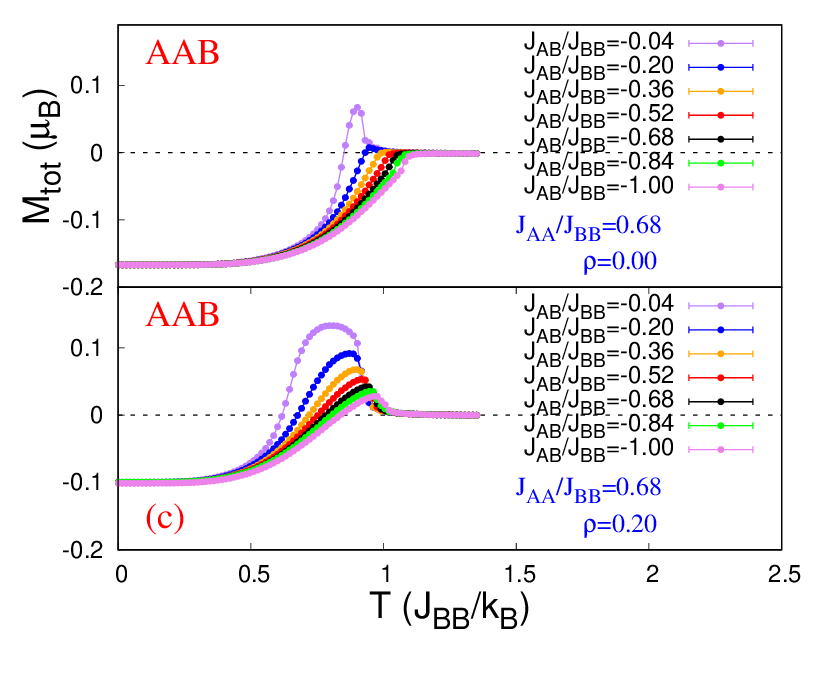}}
			\resizebox{8.75cm}{!}{\includegraphics[angle=0]{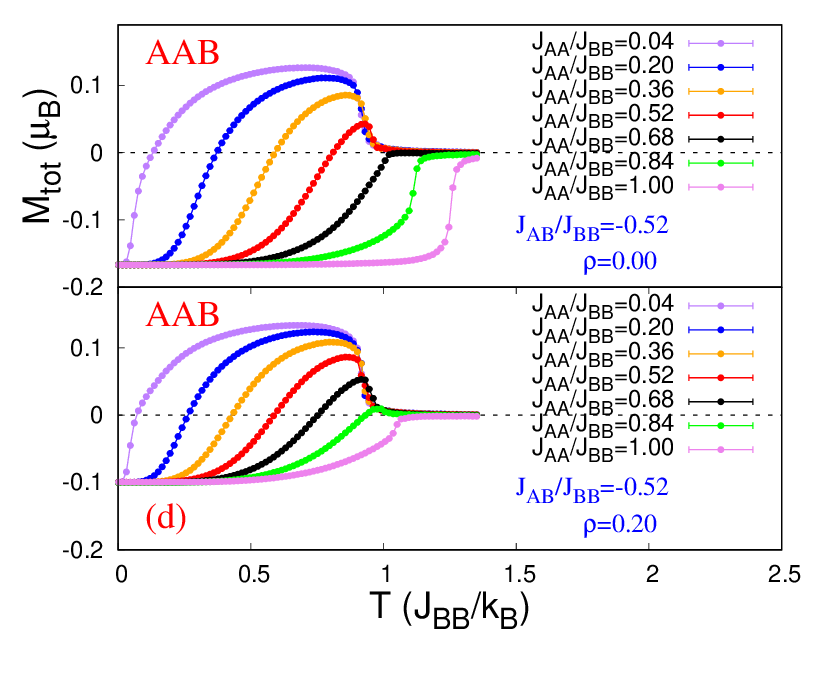}}
			
		\end{tabular}
		\caption{ (Colour Online) Plots of Order parameter, $M_{tot}$ versus dimensionless temperature, $T$, for the pure and with concentration of nonmagnetic impurities, $\rho=0.20$, for: (a) ABA: $J_{AA}/J_{BB}=0.68$ and varying $J_{AB}/J_{BB}$; (b) ABA: $J_{AB}/J_{BB}=-0.52$ and varying $J_{AA}/J_{BB}$; (c) AAB: $J_{AA}/J_{BB}=0.68$ and varying $J_{AB}/J_{BB}$; (d) AAB: $J_{AB}/J_{BB}=-0.52$ and varying $J_{AA}/J_{BB}$ . The impurity-driven creation of compensation points and shift of compensation temperatures towards the lower temperature end are visible in these cases.}
		\label{fig_2_mag_rho}
	\end{center}
\end{figure*}

It would be useful to first examine the magnetic response of the diluted configurations. In Figure \ref{fig_2_mag_rho}, a comparison has been made between the pure $(\rho=0)$ case (Courtesy \cite{Chandra3}) and with $\rho=0.20$ for both the configurations, ABA and AAB. After site-dilution, the observations are: (a) a prominent shift in the compensation temperatures and relatively small (e.g. in Figure \ref{fig_2_mag_rho}(a) with \textit{ABA}: $J_{AA}/J_{BB}=0.68$ and $J_{AB}/J_{BB}=-0.04$ and in Figure \ref{fig_2_mag_rho}(b) with \textit{ABA}: $J_{AB}/J_{BB}=-0.52$ and $J_{AA}/J_{BB}=0.04$) or undetectable (within the errorbars) shift in the critical temperatures (see Figures \ref{fig_10_crit_comp}(a) and (b)), keeping the magnetization profile same; (b) visible shift in critical temperatures leading to the creation of a compensation point, resulting in a different magnetization profile ((e.g. in Figure \ref{fig_2_mag_rho}(a) with \textit{ABA}: $J_{AA}/J_{BB}=0.68$ and $J_{AB}/J_{BB}=-0.68$ and in Figure \ref{fig_2_mag_rho}(b) with \textit{ABA}: $J_{AB}/J_{BB}=-0.52$ and $J_{AA}/J_{BB}=0.68$)).  A similar observation is valid for \textit{AAB} configuration as well [Figures \ref{fig_2_mag_rho}(c) and (d) and Figures \ref{fig_10_crit_comp}(c) and (d)].

\begin{figure*}[!htb]
	\begin{center}
		\begin{tabular}{c}
			
			\resizebox{9.0cm}{!}{\includegraphics[angle=0]{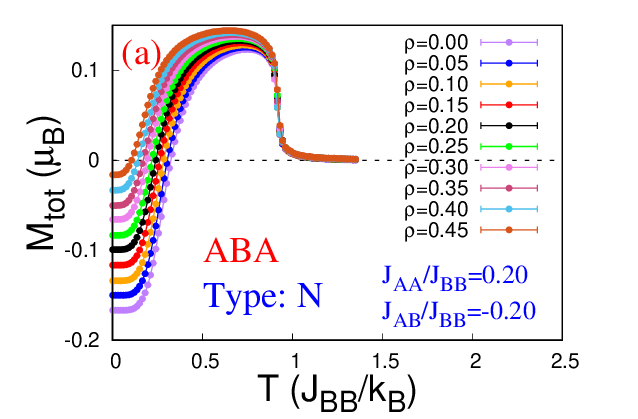}}
			\resizebox{9.0cm}{!}{\includegraphics[angle=0]{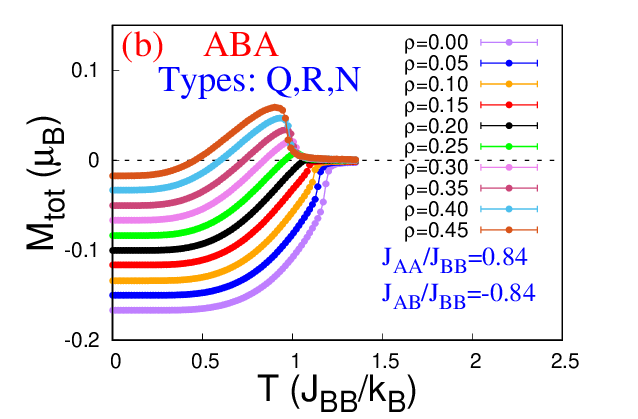}}\\
			
			\resizebox{9.0cm}{!}{\includegraphics[angle=0]{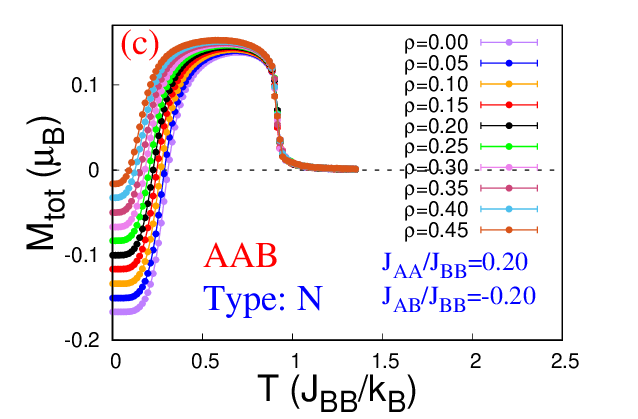}}
			\resizebox{9.0cm}{!}{\includegraphics[angle=0]{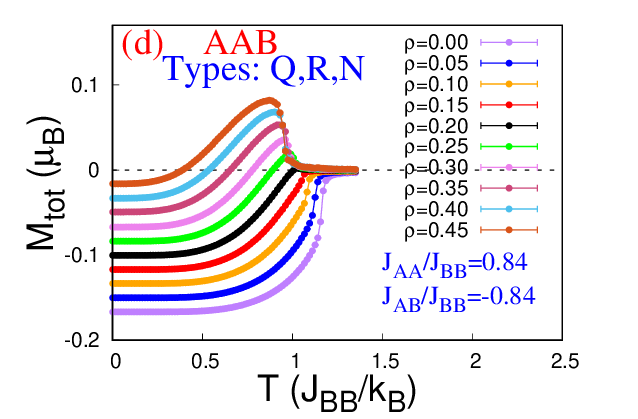}}
			
		\end{tabular}
		\caption{ (Colour Online) Plots of Order parameter, $M_{tot}$ versus dimensionless temperature, $T$, with variable concentration of nonmagnetic impurities, $\rho$, for: (a) ABA: $J_{AA}/J_{BB}=0.20$ and $J_{AB}/J_{BB}=-0.20$; (b) ABA: $J_{AA}/J_{BB}=0.84$ and $J_{AB}/J_{BB}=-0.84$; (c) AAB: $J_{AA}/J_{BB}=0.20$ and $J_{AB}/J_{BB}=-0.20$; (d) AAB: $J_{AA}/J_{BB}=0.84$ and $J_{AB}/J_{BB}=-0.84$ . Shift of the Compensation temperatures towards the lower temperature end are visible with an increase in site-dilution. The impurity-driven creation of compensation point is also witnessed for the strongest combination of coupling strengths.}
		\label{fig_3_mag_fr_afr}
	\end{center}
\end{figure*}

For both ABA and AAB types, to probe further with fixed combination of coupling strengths, the plots for the total magnetization $M_{tot}$ as a function of temperature, $T$, are shown in Figure \ref{fig_3_mag_fr_afr} with the concentration of impurities, $\rho$, acting as the parameter. We observe on increasing $\rho$, various types of magnetization profiles come into existence. According to the classification schemes of magnetization profiles, due to Neel\cite{Neel} and Strecka \cite{Strecka}: (a) Small-to-undetectable shift in critical temperatures alongwith larger shifts in compensation temperature leads to the same magnetization profile, $N$ (Ref. Figure \ref{fig_3_mag_fr_afr}(a),(c)); (b) visible shift in critical temperatures leading to the creation of compensation point, changes the magnetization profiles (Ref. Figure \ref{fig_3_mag_fr_afr}(b),(d)) from $Q$ to $R$ to $N$. The most interesting facet is the creation of compensation points in both configurations by increasing the concentration of nonmagnetic impurities for specific combinations of coupling strengths. 

\begin{figure*}[!htb]
	\begin{center}
		\begin{tabular}{c}
			
			\resizebox{9.0cm}{!}{\includegraphics[angle=0]{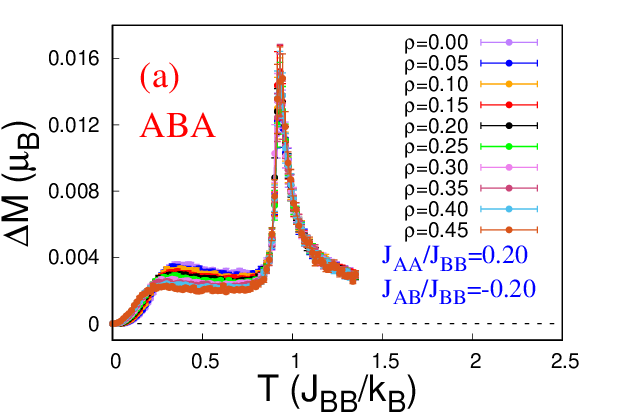}}
			\resizebox{9.0cm}{!}{\includegraphics[angle=0]{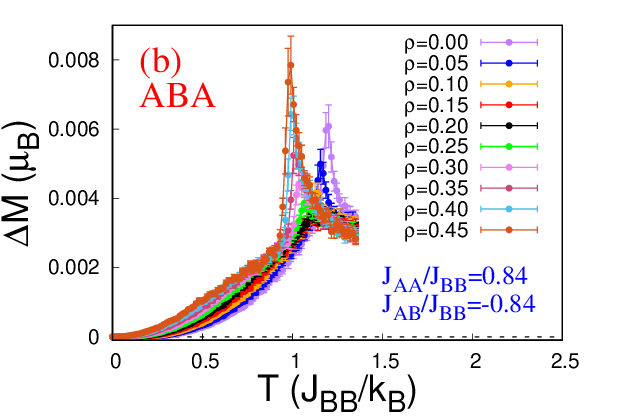}}\\
			
			\resizebox{9.0cm}{!}{\includegraphics[angle=0]{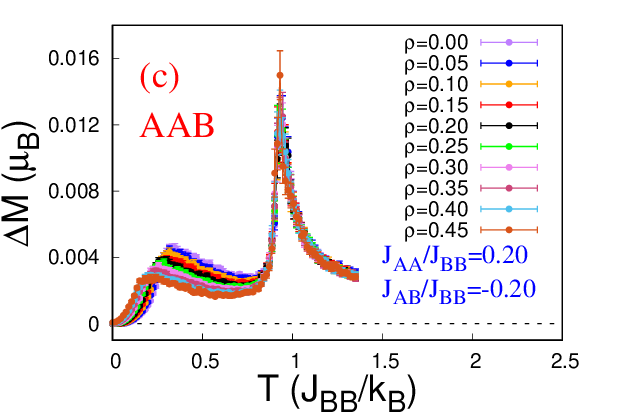}}
			\resizebox{9.0cm}{!}{\includegraphics[angle=0]{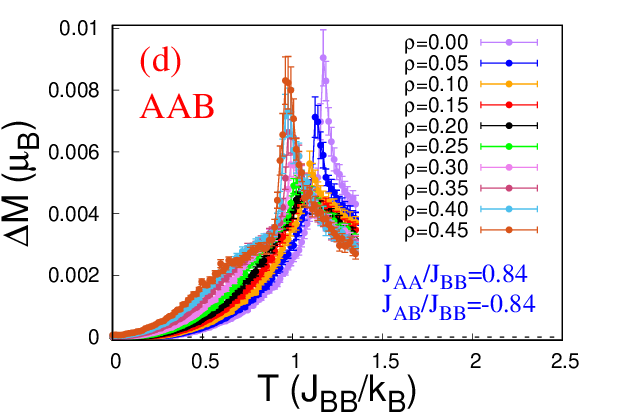}}
			
		\end{tabular}
		\caption{ (Colour Online) Plots of fluctuation of the order parameter, $\Delta M$, versus dimensionless temperature, $T$, with variable concentration of nonmagnetic impurities, $\rho$, for: (a) ABA: $J_{AA}/J_{BB}=0.20$ and $J_{AB}/J_{BB}=-0.20$; (b) ABA: $J_{AA}/J_{BB}=0.84$ and $J_{AB}/J_{BB}=-0.84$; (c) AAB: $J_{AA}/J_{BB}=0.20$ and $J_{AB}/J_{BB}=-0.20$; (d) AAB: $J_{AA}/J_{BB}=0.84$ and $J_{AB}/J_{BB}=-0.84$ .}
		\label{fig_4_flc_dm}
	\end{center}
\end{figure*}

\begin{figure*}[!htb]
	\begin{center}
		\begin{tabular}{c}
			
			\resizebox{9.0cm}{!}{\includegraphics[angle=0]{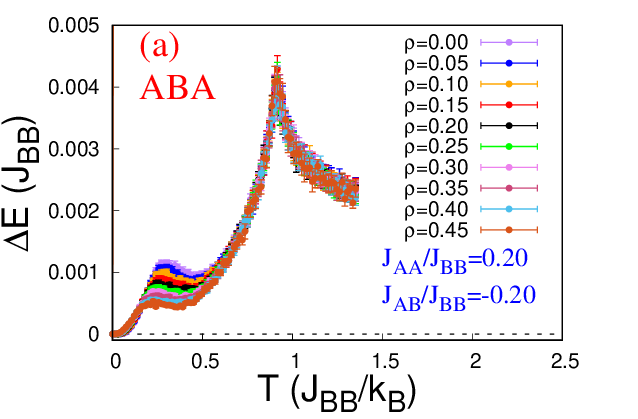}}
			\resizebox{9.0cm}{!}{\includegraphics[angle=0]{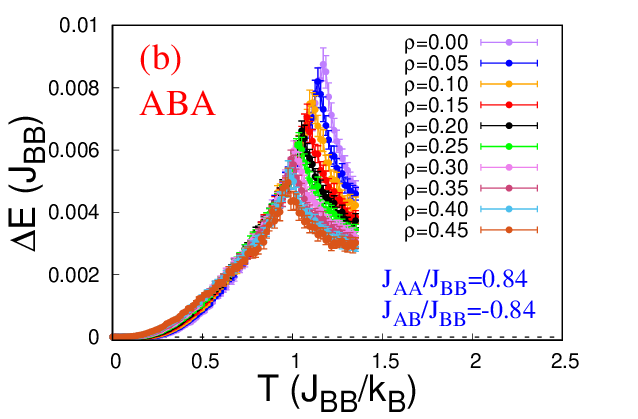}}\\
			
			\resizebox{9.0cm}{!}{\includegraphics[angle=0]{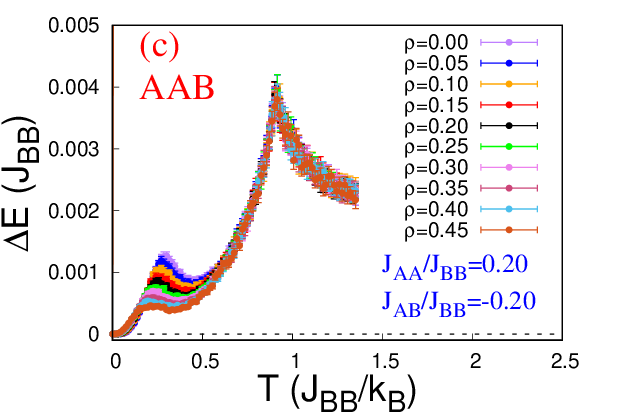}}
			\resizebox{9.0cm}{!}{\includegraphics[angle=0]{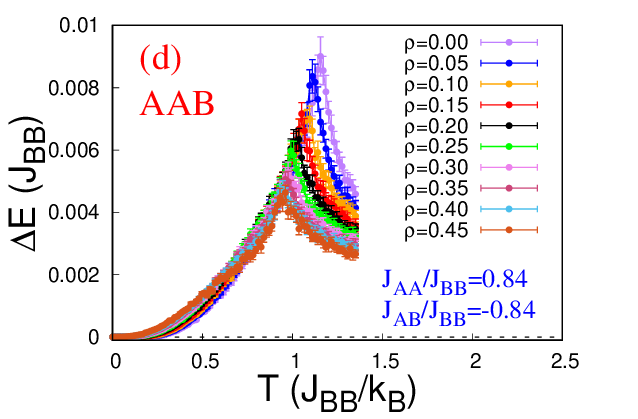}}
			
		\end{tabular}
		\caption{  (Colour Online) Plots of fluctuation of the associative energy per site, $\Delta E$, versus dimensionless temperature, $T$, with variable concentration of nonmagnetic impurities, $\rho$, for: (a) ABA: $J_{AA}/J_{BB}=0.20$ and $J_{AB}/J_{BB}=-0.20$; (b) ABA: $J_{AA}/J_{BB}=0.84$ and $J_{AB}/J_{BB}=-0.84$; (c) AAB: $J_{AA}/J_{BB}=0.20$ and $J_{AB}/J_{BB}=-0.20$; (d) AAB: $J_{AA}/J_{BB}=0.84$ and $J_{AB}/J_{BB}=-0.84$ .}
		\label{fig_5_flc_de}
	\end{center}
\end{figure*}

Now, for two specific combinations of coupling strengths, we plot both the fluctuations, of the order parameter and the associative energy per site, versus the dimensionless temperature in Figures \ref{fig_4_flc_dm} and \ref{fig_5_flc_de} respectively. The cusp-like divergence is always associated with the critical point for all the impurity concentrations for both fluctuations. Also: (a) For the coupling combination which had compensation even in the pure case, as we approach the compensation point we find the hump-like smeared peaks about it. With the increase in the impurity concentration, the dilution of bonds leads to the gradual flattening of both functions. But looking at the position of the local maximum, we understand the ordering temperature of the A-layers gradually shift towards the lower end with the increase in the impurity concentration. The position of the critical point is almost unaffected (Refer to Figures \ref{fig_4_flc_dm}(a),(c) and \ref{fig_5_flc_de}(a),(c)). (b) For the higher coupling combination, we could visually detect the maxima about the critical point shifting towards the left until a compensation point emerges. After that the behaviour is almost similar to (a). But about the compensation point, the hump-like appearance almost flattens out (Refer to Figures \ref{fig_4_flc_dm}(b),(d) and \ref{fig_5_flc_de}(b),(d)).  

\subsection{Effect of impurity}
\label{subsec_impurity}
After discussing the general results of site-dilution, we would quantitatively discuss the effects of spin-$0$ impurities in this subsection. Various compensation behaviours and magnetization profiles have technological importance \cite{Chikazumi,Miyazaki}. So it is important to find out the underlying systematics due to the site-dilution.

\begin{figure*}[!htb]
	\begin{center}
		\begin{tabular}{c}
			
			\resizebox{9.0cm}{!}{\includegraphics[angle=0]{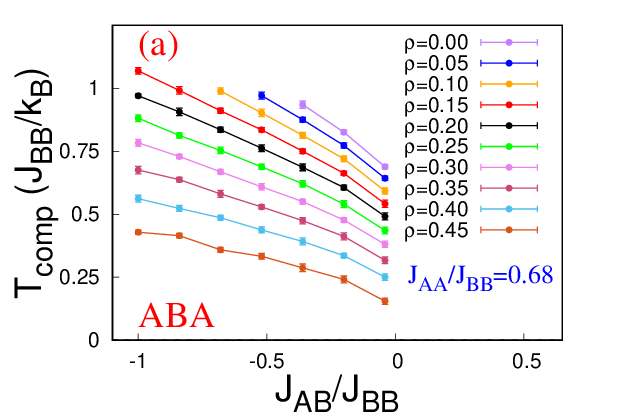}}
			\resizebox{9.0cm}{!}{\includegraphics[angle=0]{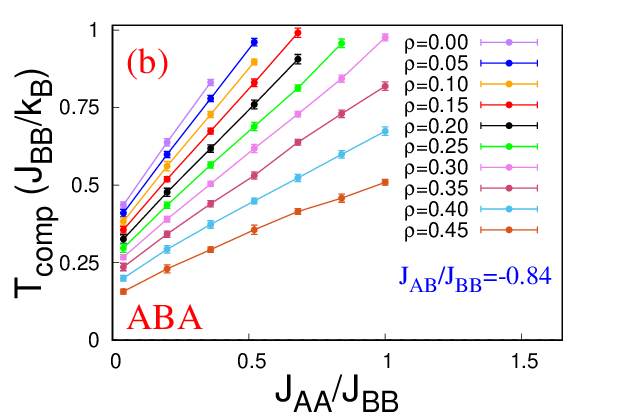}}\\
			
			\resizebox{9.0cm}{!}{\includegraphics[angle=0]{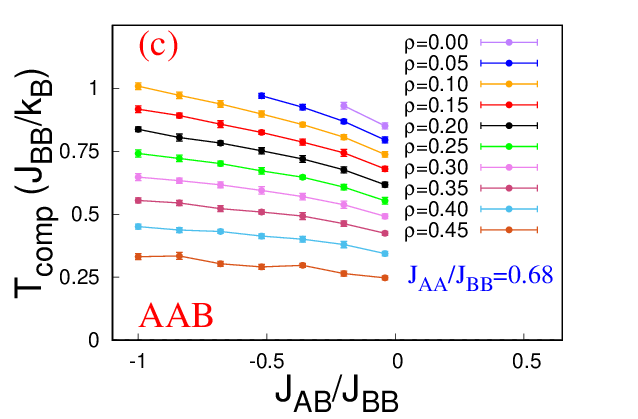}}
			\resizebox{9.0cm}{!}{\includegraphics[angle=0]{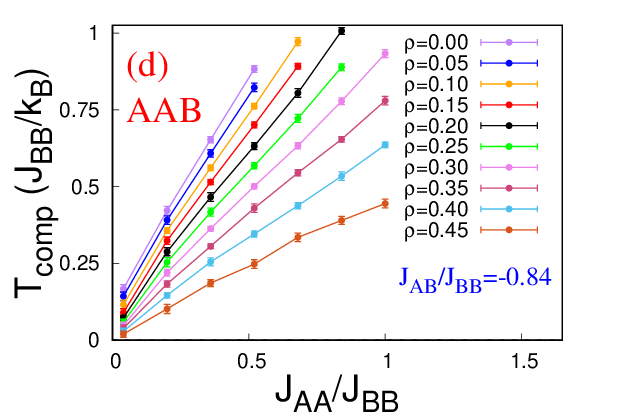}}
			
		\end{tabular}
		\caption{  (Colour Online) Plots for, ABA stacking of (a) compensation temperature, $T_{comp}$,  versus $J_{AB}/J_{BB}$, (b) compensation temperature, $T_{comp}$,  versus $J_{AA}/J_{BB}$, and AAB stacking of (c) compensation temperature, $T_{comp}$,  versus $J_{AB}/J_{BB}$, (d) compensation temperature, $T_{comp}$,  versus $J_{AA}/J_{BB}$, with variable concentration of nonmagnetic impurities, $\rho$ .}
		\label{fig_6_tcomp_rho}
	\end{center}
\end{figure*}

Figure \ref{fig_6_tcomp_rho} provides us with a general overview of what happens when we introduce the quenched disorder in the form of spin-$0$ impurities. For both configurations, as we go on increasing the concentration of spin-$0$ atoms, non-compensating magnetization profiles in the higher coupling combination regimes start to show compensation. This feature is pretty robust as keeping either of $J_{AA}/J_{BB}$ and $J_{AB}/J_{BB}$ fixed and changing the other leads to the same scenario. For example, in Figure \ref{fig_6_tcomp_rho}(a), the ABA configuration with $J_{AA}/J_{BB}=0.68$ had compensating magnetization till about $J_{AB}/J_{BB}=-0.36$ . But as we reached $\rho=0.15$, we witnessed compensating behaviours for all the antiferromagnetic coupling ratios upto $J_{AB}/J_{BB}=-1.00$. For all the other cases we witness similar kinds of behaviours, but with a different \textit{critical value of dilution}. 

\subsubsection{Phase Diagrams}
\label{subsubsec_phase}

So, the introduction of spin-$0$ impurities in the $A$-layers has a very distinct effect on the creation of compensation points. In the pure case, where compensation is absent for some parameter values, we find compensation happening in a certain diluted configuration for the same combination of parameters. What it means is the area of no-compensation shrinks as we increase the concentration of nonmagnetic impurities in the way described above. This is an important observation as after a critical concentration of spin-$0$ atoms, we witness all the combinations of coupling strengths (within the range of this study) to have compensation present.

\begin{figure*}[!htb]
	\begin{center}
		\begin{tabular}{c}
			
			\resizebox{9.0cm}{!}{\includegraphics[angle=0]{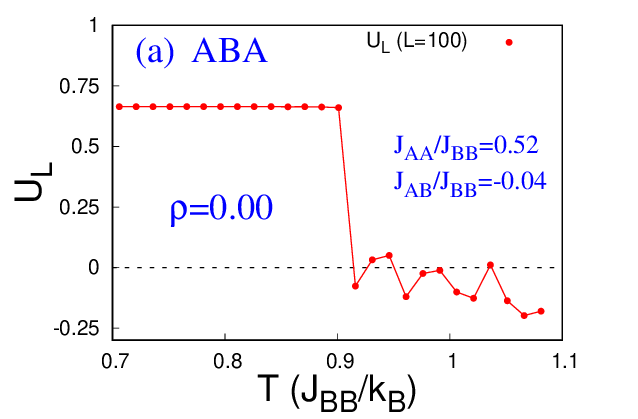}}
			\resizebox{9.0cm}{!}{\includegraphics[angle=0]{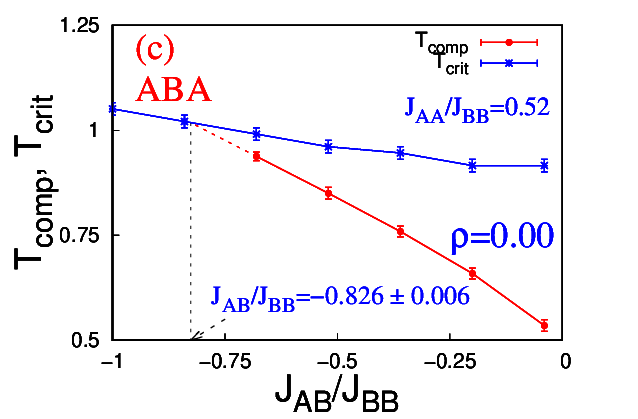}}\\
			
			\resizebox{9.0cm}{!}{\includegraphics[angle=0]{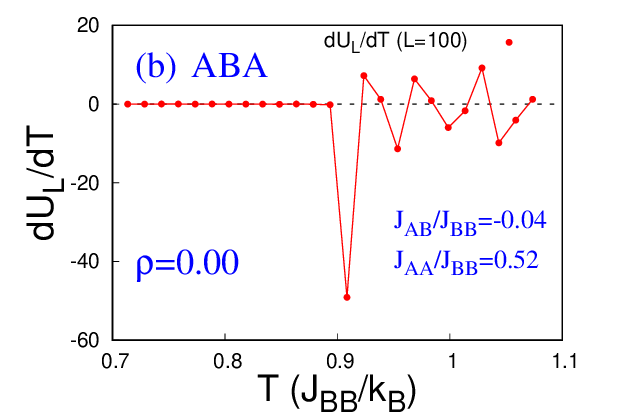}}
			\resizebox{9.0cm}{!}{\includegraphics[angle=0]{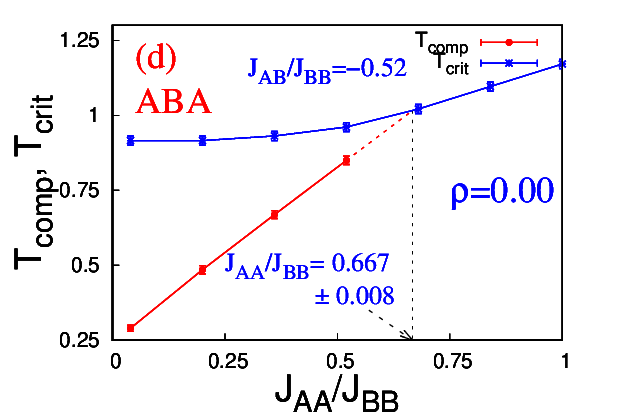}}
			
		\end{tabular}
		\caption{ (Colour Online) For a \textit{pure} ABA configuration: (a) variation of Binder's Magnetization Cumulant and (b) variation of slope of Binder's Magnetization Cumulant versus dimensionless temperature, with $J_{AA}/J_{BB}= 0.52$ and $J_{AB}/J_{BB}= -0.04$ ;and, dimensionless critical temperature $T_{crit}$ and compensation temperature $T_{comp}$ as functions of (c) $J_{AB}/J_{BB}$ with fixed $J_{AA}/J_{BB}=0.52$ (d) $J_{AA}/J_{BB}$ with fixed $J_{AB}/J_{BB}=-0.52$. The intersections of vertical dashed lines with the horizontal axis, mark the value of either $J_{AB}/J_{BB}$ or $J_{AA}/J_{BB}$, below/above which no compensation is detected. Where the errorbars are not visible, they are smaller than the point markers.}
		\label{fig_7_bincum_intersec}
	\end{center}
\end{figure*}

The procedure of detecting the critical temperature for a specific combination of coupling strengths is outlined in Figures \ref{fig_7_bincum_intersec}(a) and (b). From the initial temperature series of Binder's magnetization cumulant, $U_{L}$, for the system size, $3L^{2}=3\times 100 \times 100$,  we would plot the temperature derivative of $U_{L}$ i.e. $\dfrac{dU_{L}}{dT}$ versus the dimensionless temperature. At the \textit{pseudo-critical} point, $\dfrac{dU_{L}}{dT}$ attains minima, when the temperature is increased as in Figure \ref{fig_7_bincum_intersec}(b) \cite{Ferrenberg}. Next up, to draw the phase diagram in parameter space, we need to detect the combinations of coupling strengths across the entire parameter space. Two examples are shown in Figures \ref{fig_7_bincum_intersec}(c) and (d). After keeping either the ferromagnetic or anti-ferromagnetic coupling ratio fixed, we plot the compensation temperature, $T_{comp}$ and critical temperature, $T_{crit}$ as a function of the other variable interaction strength. At the lower end, for fixed ferromagnetic strength, we would linearly extrapolate the curve for $T_{comp}$ to find out the intersection with the curve for $T_{crit}$. The x-coordinate of the intersecting point is the minimum value of antiferromagnetic strength where compensation ceases to exist [Refer to Figure \ref{fig_7_bincum_intersec}(c)]. For fixed anti-ferromagnetic strength, the intersection finding is to be done at the higher end to find out the maximum value of ferromagnetic strength [Refer to Figure \ref{fig_7_bincum_intersec}(d)]. The upper bounds of linear interpolation procedure provide the errors associated with the intersections \cite{Scarborough}.

\begin{figure*}[!htb]
	\begin{center}
		\begin{tabular}{c}
			
			\resizebox{9.75cm}{!}{\includegraphics[angle=-90]{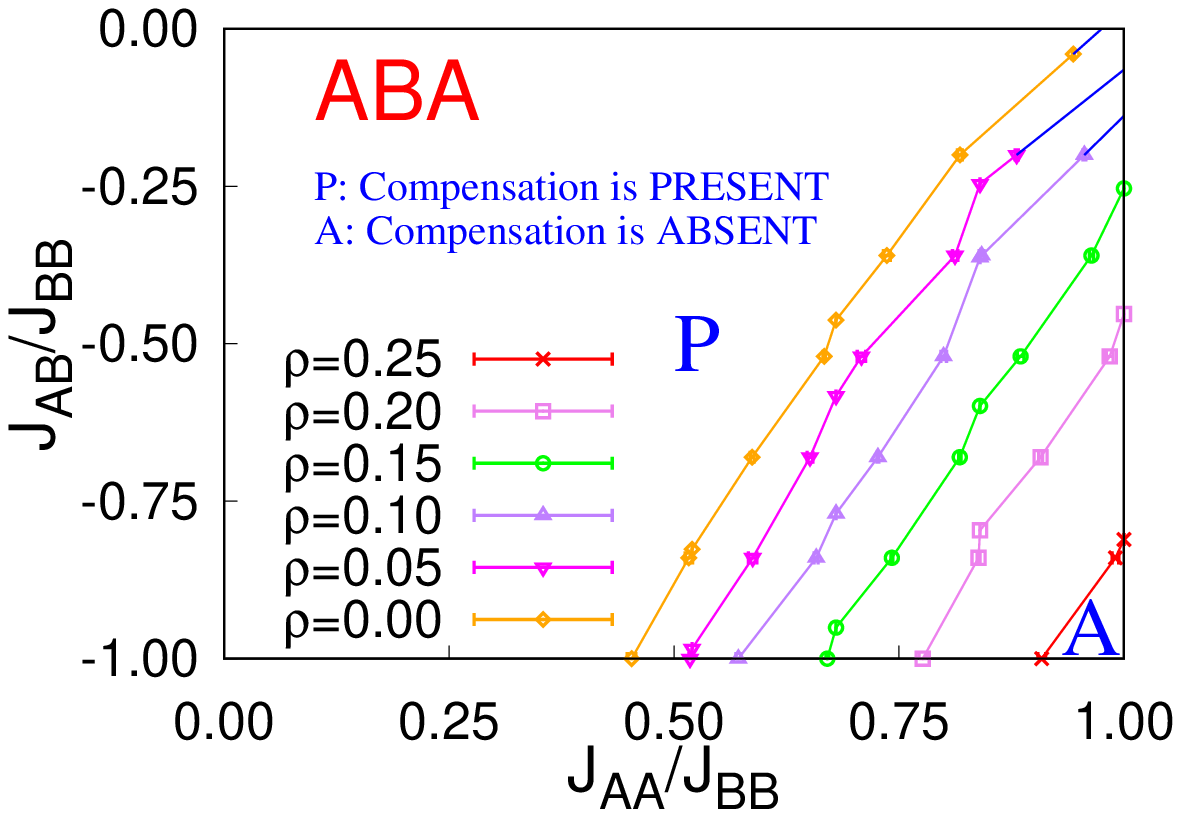}}
			\resizebox{9.75cm}{!}{\includegraphics[angle=-90]{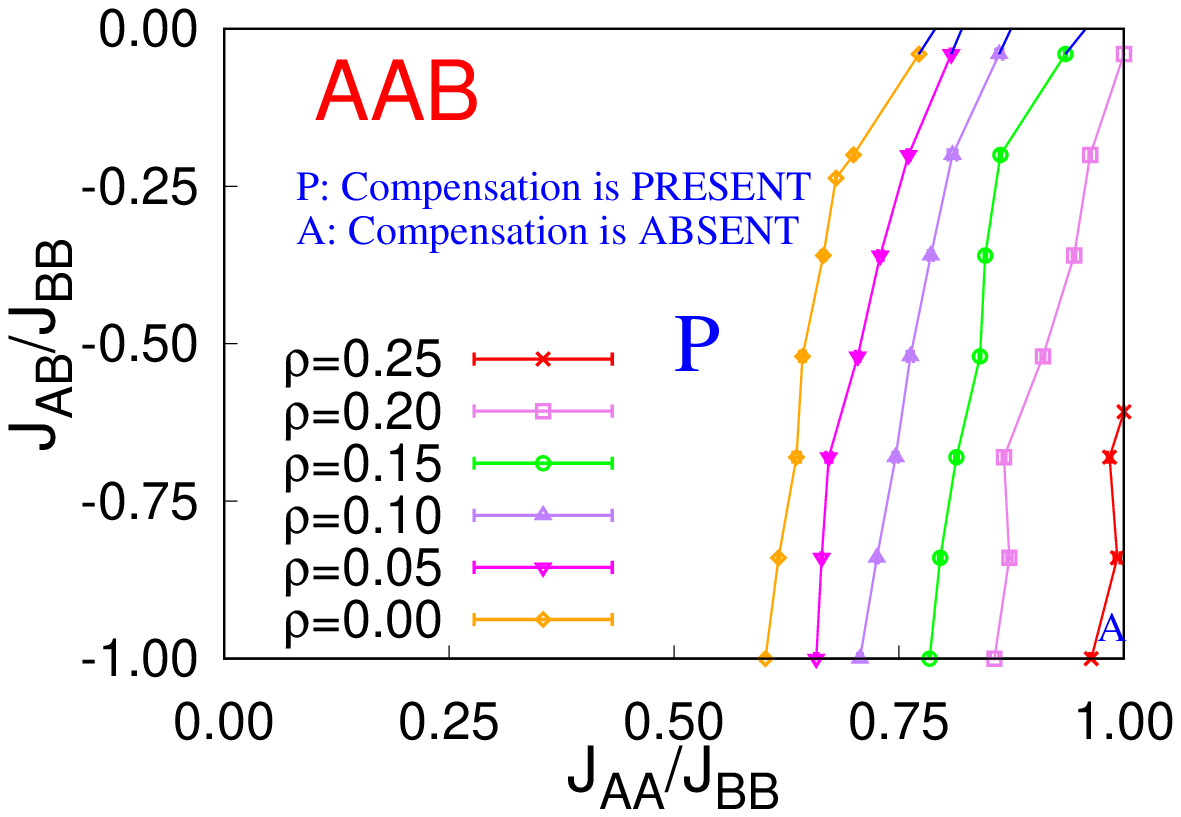}}
			
		\end{tabular}
		\caption{ (Colour Online) Phase diagrams in the $\left(J_{AB}/J_{BB} \times J_{AA}/J_{BB}\right)$ plane for: (a) ABA and (b) AAB configurations with concentration of nonmagnetic atoms acting as a parameter. The impurity-driven changes in the relative areas in the Phase diagram are clearly visible. The blue ends of the curves are obtained via linear extrapolation.}
		\label{fig_8_phase_diagram}
	\end{center}
\end{figure*}

\begin{figure*}[!htb]
	\begin{center}
		\begin{tabular}{c}
			
			\resizebox{9.75cm}{!}{\includegraphics[angle=0]{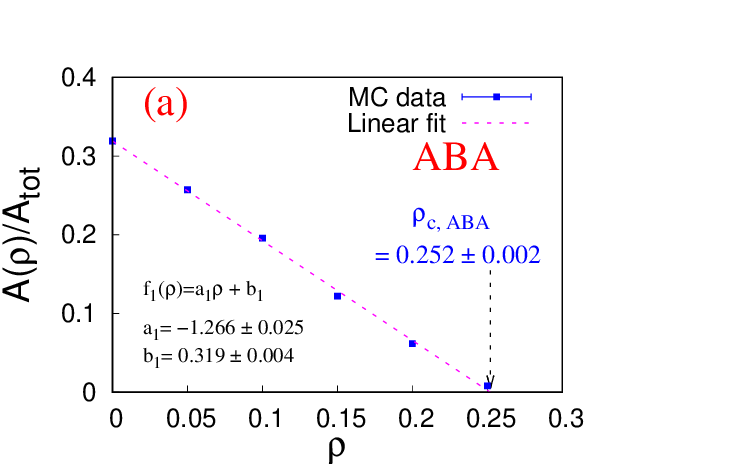}}
			\resizebox{9.75cm}{!}{\includegraphics[angle=0]{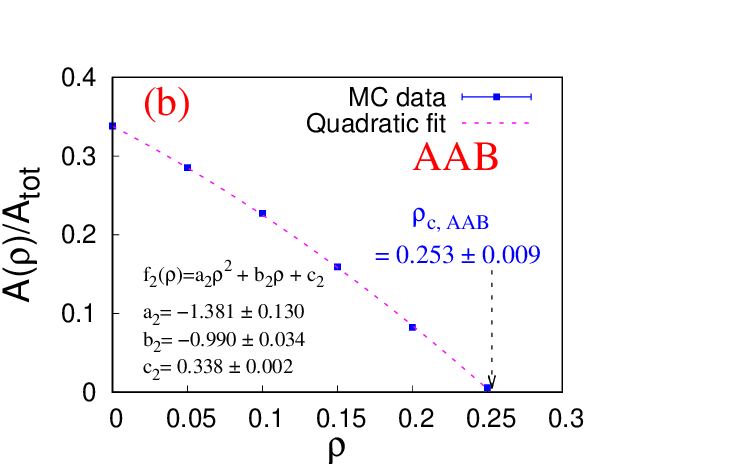}}
			
		\end{tabular}
		\caption{ (Colour Online) Plots of the fractional area of the compensating region in the $\left(J_{AB}/J_{BB} \times J_{AA}/J_{BB}\right)$ plane, $A(\rho)/A_{tot}$ versus concentration of the nonmagnetic impurities, $\rho$ for the: (a) ABA and (b) AAB configurations with a faithful fitting function.}
		\label{fig_9_phase_area_fit}
	\end{center}
\end{figure*}

In Figure \ref{fig_8_phase_diagram}, we plot the phase diagrams in the parameter space ($J_{AB}/J_{BB}$ $\times$ $J_{AA}/J_{BB}$) for both the ABA and AAB configurations. The gradual diminishing of the A-marked areas (absence of compensation) as we increase the concentration of dilution leads us to mathematically fit the $A(\rho)/A_{tot}$ versus $\rho$ data in Figure \ref{fig_9_phase_area_fit}. To find out the values of $A(\rho)$, for different $\rho$, the integration is performed by the Monte Carlo method \cite{Krauth}. In a similar situation, this method is previously applied in \cite{Chandra4,Chandra5} where technical details can be revisited.

For ABA, the linear fit is adequate with the fitting polynomial: 
\begin{equation}
\dfrac{A(\rho)}{A_{tot}} = a_{1}\rho + b_{1}
\end{equation}
where, $a_{1}=-1.266 \pm 0.025$ and $b_{1}= 0.319 \pm 0.004$ . The critical concentration of ABA configuration is: $\rho_{c,ABA}=0.252 \pm 0.002$ .

For AAB, the quadratic fit is adequate with the fitting polynomial: 
\begin{equation}
\dfrac{A(\rho)}{A_{tot}} = a_{2}\rho^{2} + b_{2} \rho + c_{2}
\end{equation}
where, $a_{2}=-1.381 \pm 0.130$, $b_{2}= -0.990 \pm 0.034$ and $c_{2}= 0.338 \pm 0.002$. The critical concentration of AAB configuration is: $\rho_{c,AAB}=0.253 \pm 0.009$ .

\subsubsection{Systematics}
\label{subsubsec_systematics}

\begin{figure*}[!htb]
	\begin{center}
		\begin{tabular}{c}
			
			\resizebox{9.0cm}{!}{\includegraphics[angle=0]{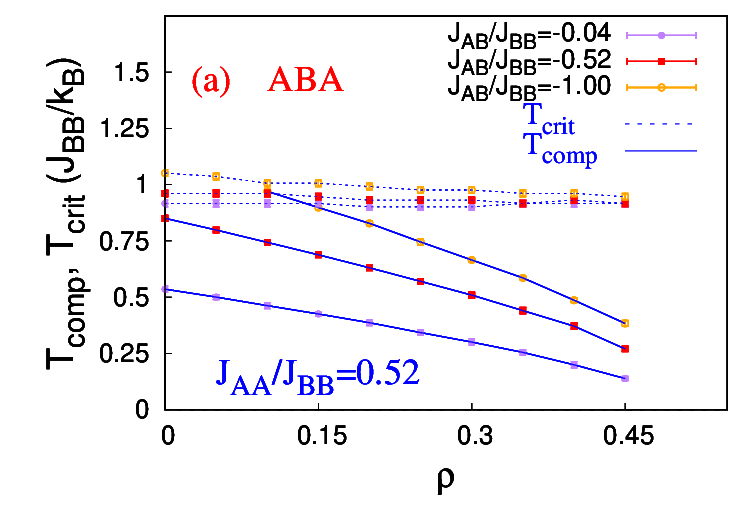}}
			\resizebox{9.0cm}{!}{\includegraphics[angle=0]{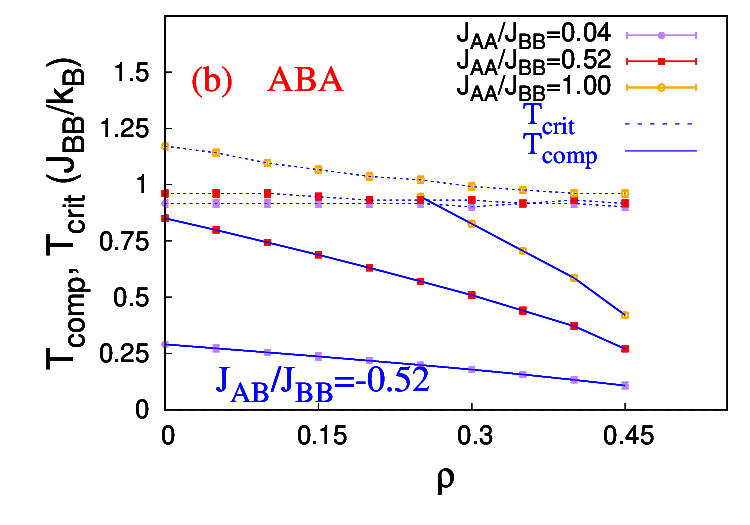}}\\
			
			\resizebox{9.0cm}{!}{\includegraphics[angle=0]{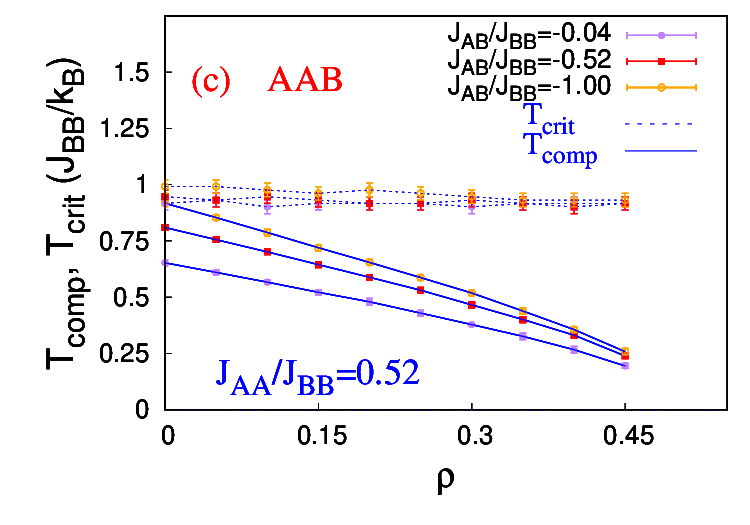}}
			\resizebox{9.0cm}{!}{\includegraphics[angle=0]{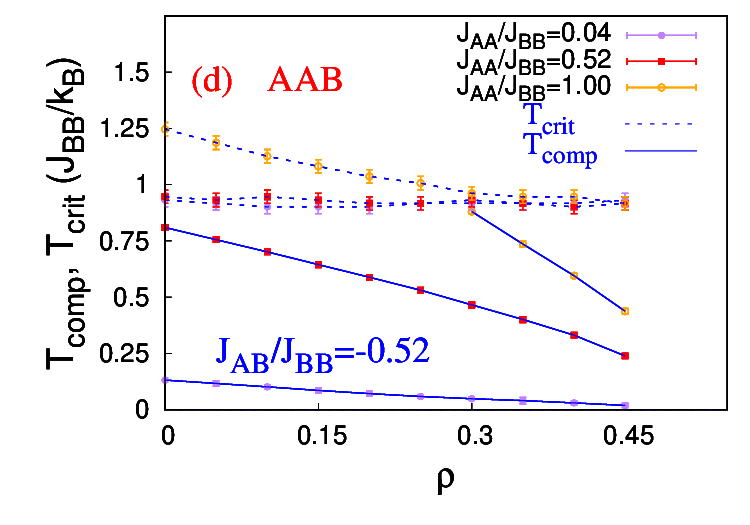}}
			
		\end{tabular}
		\caption{ (Colour Online) Variation of Critical and Compensation temperatures with an increase in the concentration of nonmagnetic impurities for a few representative cases: for an ABA configuration in (a) and (b) and for an AAB configuration in (c) and (d). Where the errorbars are not visible, they are smaller than the point markers.}
		\label{fig_10_crit_comp}
	\end{center}
\end{figure*}

In this section, we would try to theoretically characterize the diluted ferrimagnetic systems. In Figure \ref{fig_10_crit_comp}, we witness the variation of critical and compensation temperatures with the concentration of spin-$0$ atoms for a few combinations of coupling strengths. For both the configurations, ABA and AAB, we observe the compensation temperatures to vary in a \textit{nonlinear fashion} as we increase site-dilution. Alongside, the change in critical temperatures is limited to within $\sim23\%$ in ABA and $\sim26\%$ in AAB till $45\%$ dilution. These features make such magnetic heterostructures very interesting. Next, we would like to model the variation of the compensation temperatures as the $N$-type magnetization profiles (having a compensation point) find many important technological applications \cite{Chikazumi,Miyazaki}. 

\begin{figure*}[!htb]
 	\begin{center}
 		\begin{tabular}{c}
 			
 			\resizebox{9.0cm}{!}{\includegraphics[angle=0]{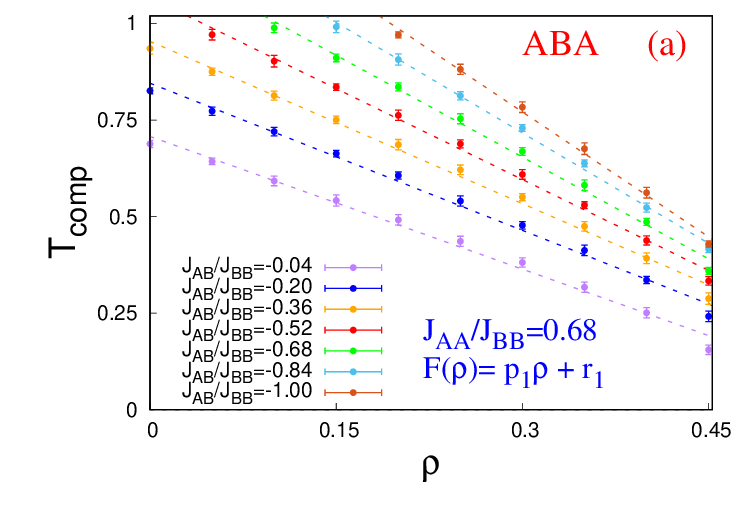}}
 			\resizebox{9.0cm}{!}{\includegraphics[angle=0]{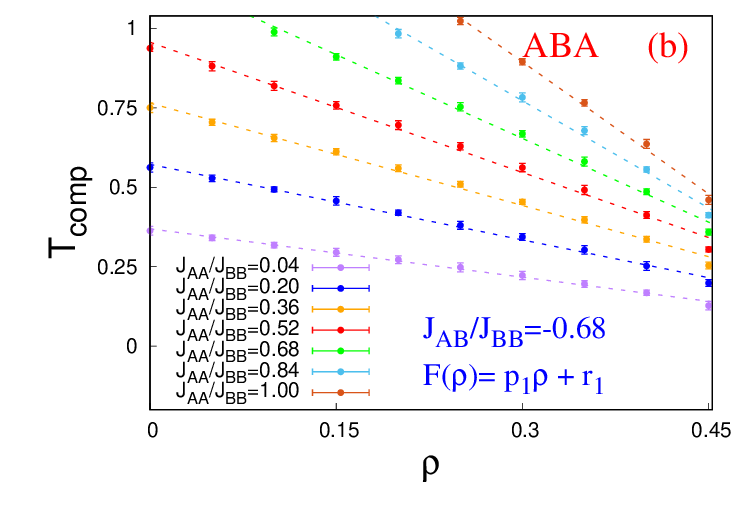}}\\
 			
 			\resizebox{9.0cm}{!}{\includegraphics[angle=0]{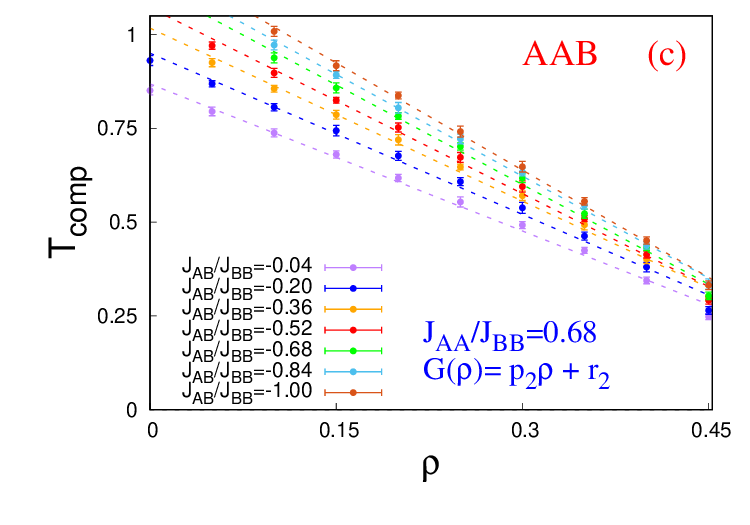}}
 			\resizebox{9.0cm}{!}{\includegraphics[angle=0]{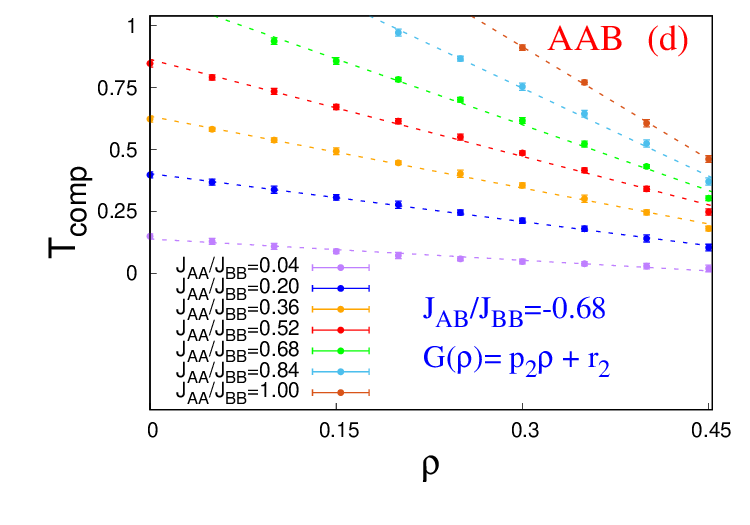}}
 			
 		\end{tabular}
 		\caption{ (Colour Online) Fitting of Compensation temperatures with the increase in the concentration of nonmagnetic impurities for a few representative cases: for an ABA configuration in (a) and (b) and for an AAB configuration in (c) and (d). Where the errorbars are not visible, they are smaller than the point markers.}
 		\label{fig_11_comp_fit}
 	\end{center}
 \end{figure*}

Now to model the variation of compensation temperatures with concentration of nonmagnetic impurities, a linear polynomial may be employed. The mathematical forms follow:\\
For ABA:
\begin{equation}
\label{eq_tcomp_fit_ABA}
T_{comp,ABA}(\rho,J_{AA}/J_{BB},J_{AB}/J_{BB})= p_{1} \rho + r_{1}
\end{equation}
For AAB:
\begin{equation}
\label{eq_tcomp_fit_AAB}
T_{comp,AAB}(\rho,J_{AA}/J_{BB},J_{AB}/J_{BB})= p_{2} \rho + r_{2}
\end{equation}
The explicit dependence of the coupling strengths is contained in the coefficients at the \textit{right-hand side} of the set of equations. That means, $p_{i}\equiv p_{i}(J_{AA}/J_{BB},J_{AB}/J_{BB})$ ; $r_{i}\equiv r_{i}(J_{AA}/J_{BB},J_{AB}/J_{BB})$ with $i\in \{1,2\}$ . These particular choices of fitting formulae provide us with \textit{reduced chisquare}, $\chi^{2}/n_{DOF} \sim 1$ and the asymptotic errors in the fitting parameters are \textit{mostly confined} below $10\%$ . So, this choice of mathematical forms is reliable for further calculations. A few selective examples are provided in Figure \ref{fig_11_comp_fit} for both configurations. A discussion on the fitting parameters is necessary \cite{Press} and in Appendix \ref{appendix_fit}, we show the behaviours of the fitting parameters for both configurations.

\begin{figure*}[!htb]
	\begin{center}
		\begin{tabular}{c}
			
			(a)\resizebox{8.50cm}{!}{\includegraphics[angle=0]{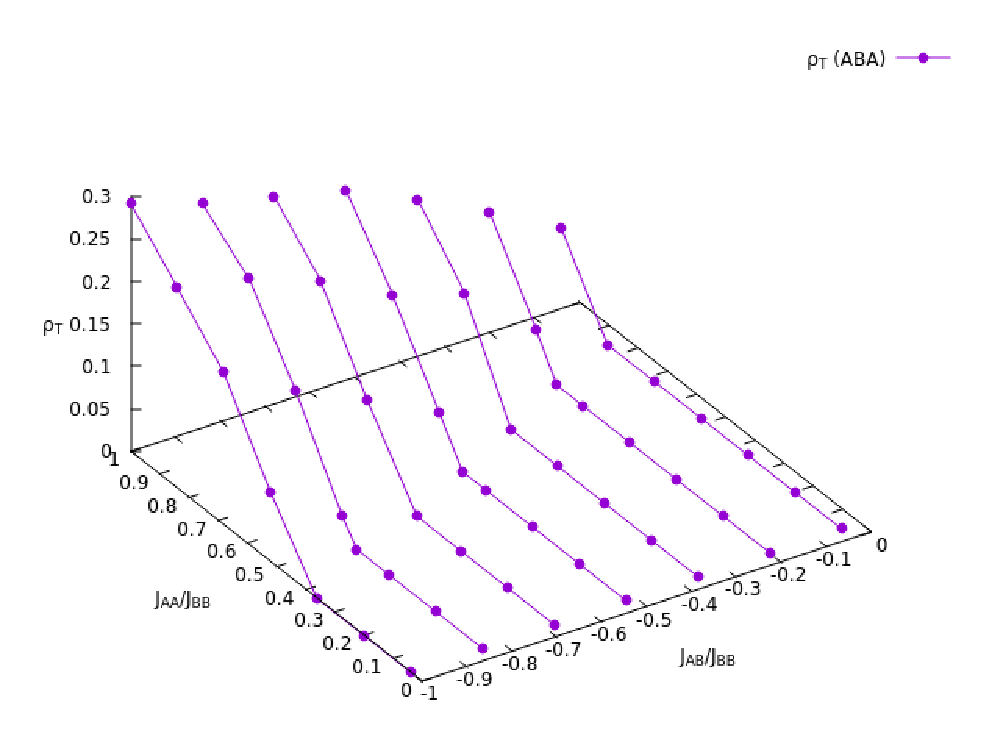}}
			(b)\resizebox{8.50cm}{!}{\includegraphics[angle=0]{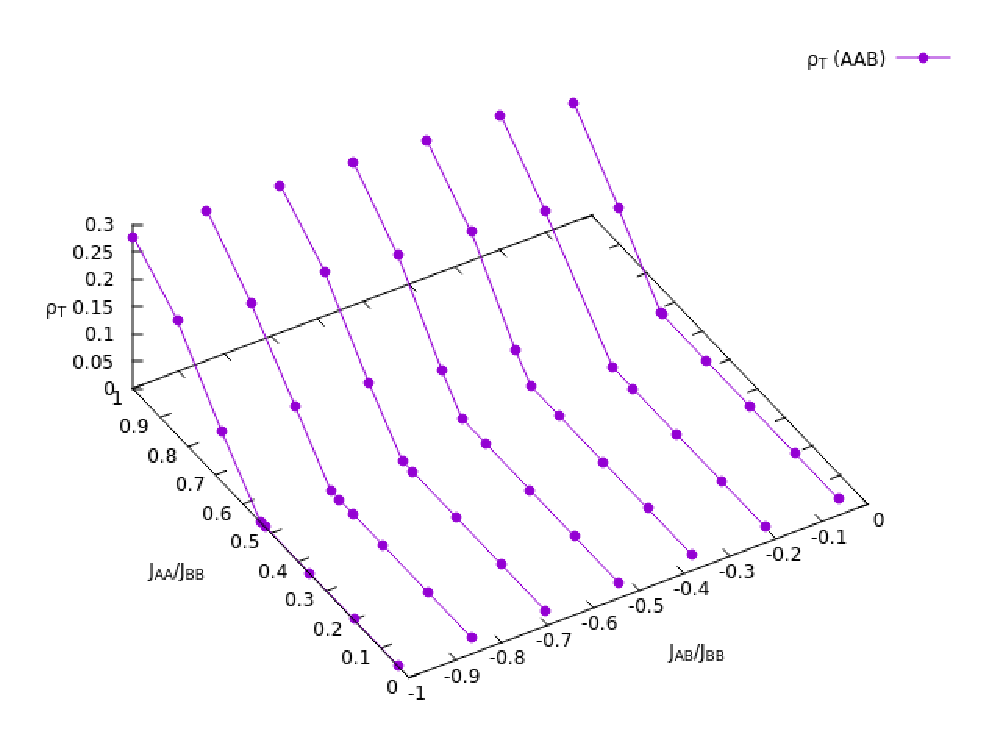}}
			
		\end{tabular}
		\caption{ (Colour Online) Plots of threshold concentration of spin-$0$ impurities versus Hamiltonian parameters, $J_{AA}/J_{BB}$ and $J_{AB}/J_{BB}$ : for an ABA configuration in (a) and for an AAB configuration in (b) .}
		\label{fig_12_thresconc_3d}
	\end{center}
\end{figure*}

Now, we would find out the threshold value of the concentration of nonmagnetic impurities for any given combination of coupling strengths, above which we would witness compensation. First, in Figure \ref{fig_12_thresconc_3d}, we would witness how the threshold impurity concentrations vary with two Hamiltonian parameters, $J_{AA}/J_{BB}$ and $J_{AB}/J_{BB}$ . It is evident that as the magnitude of either of the Coupling ratios increases, the values of the threshold concentration of nonmagnetic impurities, $\rho_{T}$, increase. This motivates us to find out how $\rho_{T}$ behaves as a function of $J_{AA}/J_{BB}$ and $J_{AB}/J_{BB}$ . Essentially, for each configuration, we will then have two 2D plots. 

 \begin{figure*}[!htb]
	\begin{center}
		\begin{tabular}{c}
			
			(a)\resizebox{8.50cm}{!}{\includegraphics[angle=0]{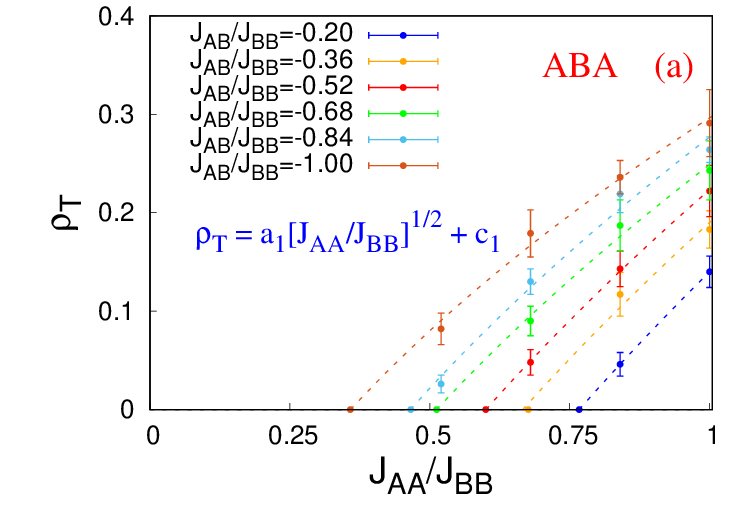}}
			(b)\resizebox{8.50cm}{!}{\includegraphics[angle=0]{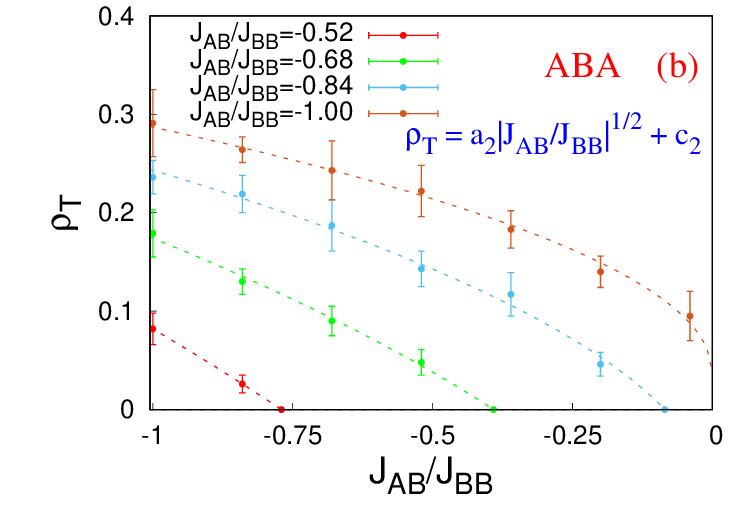}}
			
		\end{tabular}
		\caption{ (Colour Online) Plots of threshold concentration of spin-$0$ impurities versus Hamiltonian parameters: (a) $\rho_{T}$ versus $J_{AA}/J_{BB}$ and (b) $\rho_{T}$ versus $J_{AB}/J_{BB}$; for an ABA configuration.}
		\label{fig_13_aba_thresconc_2d}
	\end{center}
\end{figure*}

For the ABA configuration, from Figure \ref{fig_13_aba_thresconc_2d}, we see a systematic variation of $\rho_{T}$ . A mathematical fit now can be attempted:
\begin{equation}
\label{eq_psi1}
\Psi_{1}(J_{AA}/J_{BB}, J_{AB}/J_{BB})=a_{1}\sqrt{J_{AA}/J_{BB}} + c_{1}
\end{equation}
for fixed $J_{AB}/J_{BB}$ [Refer to Figure \ref{fig_13_aba_thresconc_2d}(a)]. And,

\begin{equation}
\label{eq_psi2}
\Psi_{2}(J_{AA}/J_{BB}, J_{AB}/J_{BB})=a_{2}\sqrt{|J_{AB}/J_{BB}|} + c_{2}
\end{equation}
for fixed $J_{AA}/J_{BB}$ [Refer to Figure \ref{fig_13_aba_thresconc_2d}(b)].
Along with, the coefficients are functions of coupling ratios e.g. $a_{1}\equiv a_{1}(J_{AB}/J_{BB})$, $c_{1}\equiv c_{1}(J_{AB}/J_{BB})$ and $a_{2}\equiv a_{2}(J_{AA}/J_{BB})$, $c_{2}\equiv c_{2}(J_{AA}/J_{BB})$ .

\begin{figure*}[!htb]
	\begin{center}
		\begin{tabular}{c}
			
			(a)\resizebox{8.50cm}{!}{\includegraphics[angle=0]{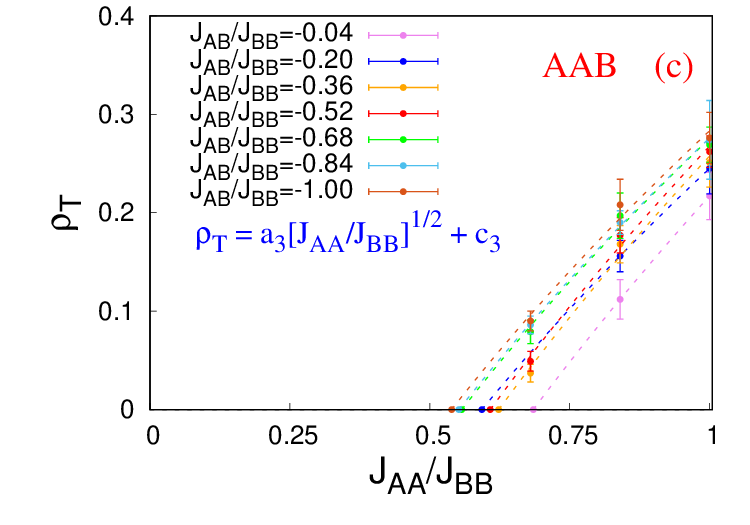}}
			(b)\resizebox{8.50cm}{!}{\includegraphics[angle=0]{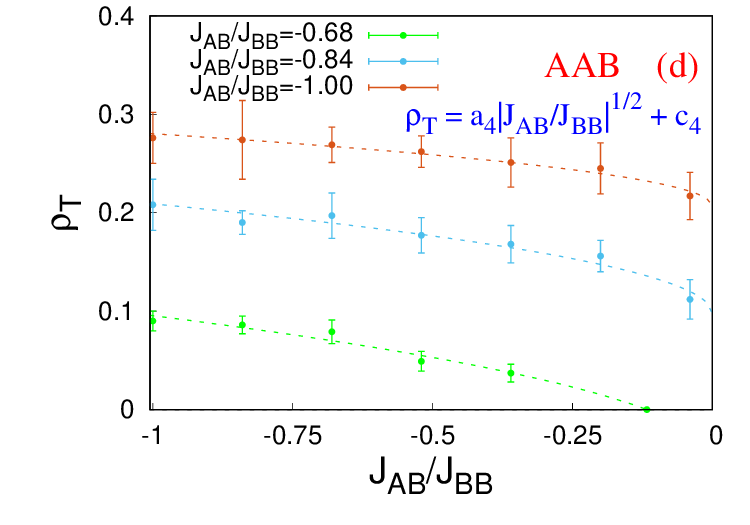}}
			
		\end{tabular}
		\caption{ (Colour Online) Plots of threshold concentration of spin-$0$ impurities versus Hamiltonian parameters: (a) $\rho_{T}$ versus $J_{AA}/J_{BB}$ and (b) $\rho_{T}$ versus $J_{AB}/J_{BB}$; for an AAB configuration.}
		\label{fig_14_aab_thresconc_2d}
	\end{center}
\end{figure*}

For the AAB configuration, from Figure \ref{fig_14_aab_thresconc_2d}, we see a similar variation of $\rho_{T}$ . Thus similar mathematical forms can be proposed:
\begin{equation}
\label{eq_phi1}
\Phi_{1}(J_{AA}/J_{BB}, J_{AB}/J_{BB})=a_{3}\sqrt{J_{AA}/J_{BB}} + c_{3}
\end{equation}
for fixed $J_{AB}/J_{BB}$ [Refer to Figure \ref{fig_14_aab_thresconc_2d}(a)]. And,

\begin{equation}
\label{eq_phi2}
\Phi_{2}(J_{AA}/J_{BB}, J_{AB}/J_{BB})=a_{4}\sqrt{|J_{AB}/J_{BB}|} + c_{4}
\end{equation}
for fixed $J_{AA}/J_{BB}$ [Refer to Figure \ref{fig_14_aab_thresconc_2d}(b)].
Along with, the coefficients are functions of coupling ratios e.g. $a_{3}\equiv a_{3}(J_{AB}/J_{BB})$, $c_{3}\equiv c_{3}(J_{AB}/J_{BB})$ and $a_{4}\equiv a_{4}(J_{AA}/J_{BB})$, $c_{4}\equiv c_{4}(J_{AA}/J_{BB})$ .

The value of threshold concentration of spin-$0$ impurities at any of the intersecting points on the Hamiltonian parameter space can be obtained through the geometric mean of the two perpendicular approaches via the functions presented in Equations \ref{eq_psi1} to \ref{eq_phi2}: one along the $J_{AA}/J_{BB}$ axis and the other along the $J_{AB}/J_{BB}$ axis.  

\section{Summary}
\label{sec_summary}
In this work, we have investigated the role played by site dilution in creating compensation points in a ferrimagnetic spin-1/2 Ising trilayer on triangular monolayers compared to its pristine counterpart, using Metropolis Monte Carlo simulation. The ABA and AAB configurations have non-equivalent planes in the sense that only A-layers are randomly site-diluted but the B-layer, \textit{with dominant in-plane coupling}, is pristine. While discussing the magneto-thermal behaviour, we witnessed that for a fixed concentration of spin-$0$ impurities, the compensation point shifts towards the critical point and ultimately merges with it, as we increase either of the coupling ratios. For a fixed combination of coupling ratios, an increase in the concentration of the diluted sites leads to the shift of both, compensation and critical temperatures towards the lower temperature ends. Thermal variation of fluctuations of magnetisation and associative energy also supports the previous observation. The variation of Compensation temperature with the concentration of nonmagnetic atoms can be modelled by a \textit{linear} variation. Across the entire range of concentration of nonmagnetic impurities, we witness continuous phase transitions (second order phase transitions) across the critical points for all the combinations of coupling strengths. Still, the most interesting and important observation is the impurity-driven creation of compensation points for certain combinations of coupling strengths in site-diluted systems. The threshold concentration of nonmagnetic impurities, $\rho_{T}$, above which a compensation point is created for a certain combination of coupling ratios in a diluted trilayered system, varies in a parabolic manner with the coupling ratios for both the ABA and AAB configurations. Thus the fitted mathematical formulae characterise the system and provide us with a complete description of the diluted trilayered magnetic systems. These results may contribute to the prospective research of diluted magnetic layered materials with non-equivalent planes with several different sublattice geometries.

\section*{Acknowledgements}
The author gratefully acknowledges financial assistance from the University Grants Commission, India in the form of a Research fellowship and extends his thanks to Dr. Debabrata Ghorai and Dr. Tamaghna Maitra for providing the computational facilities. Insightful comments and suggestions by the anonymous referees are also gratefully acknowledged.

\newpage
\begin{center} {\Large \textbf {References}} \end{center}

\vskip 2cm
\appendix
\begin{flushleft}
	{\LARGE \textbf{Appendix}}
\end{flushleft}
\section{Behaviour of the fitting parameters}
\label{appendix_fit}
After mathematically fitting the behaviours of the Compensation temperatures with dilution in Section \ref{subsubsec_systematics}, here we will discuss how the fitting parameters behave as we vary the independent variables. After finding a reliable fitting function, if the set of parameters turns out to be well-behaved, the description then becomes indeed a good one. 

\begin{figure*}[!htb]
	\begin{center}
		\begin{tabular}{c}
			
			\resizebox{8.0cm}{!}{\includegraphics[angle=0]{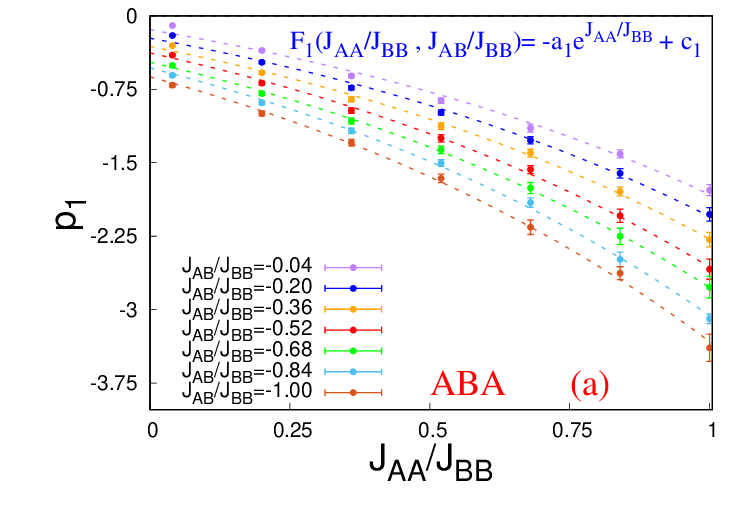}}
			\resizebox{8.0cm}{!}{\includegraphics[angle=0]{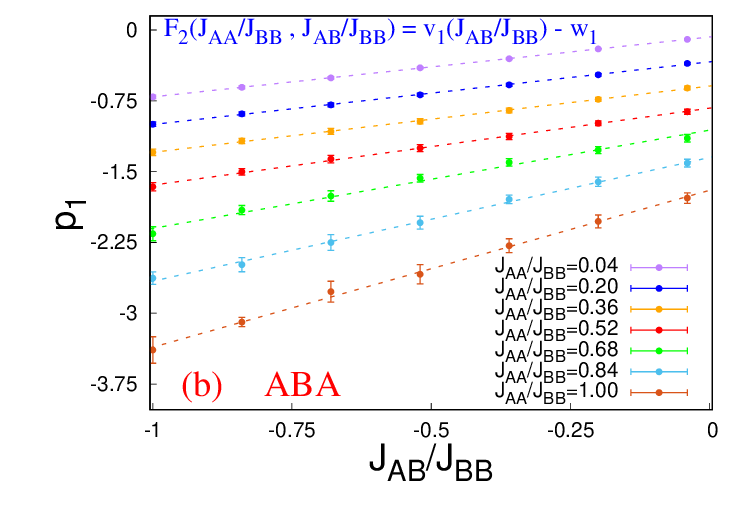}}\\
			
			\resizebox{8.0cm}{!}{\includegraphics[angle=0]{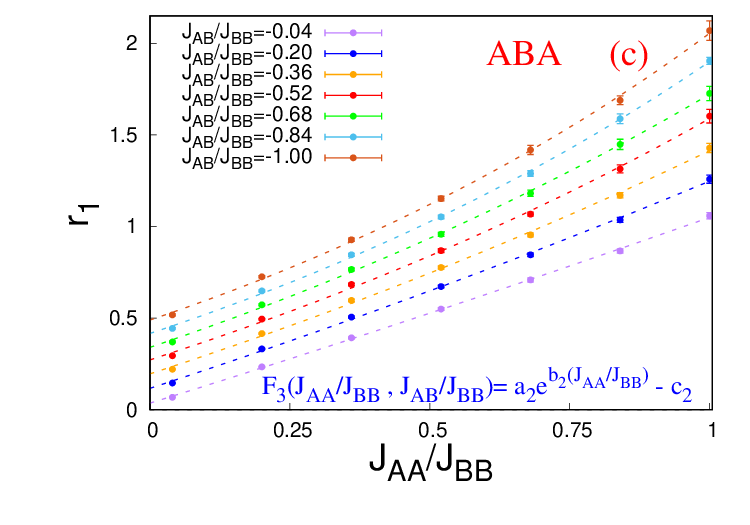}}
			\resizebox{8.0cm}{!}{\includegraphics[angle=0]{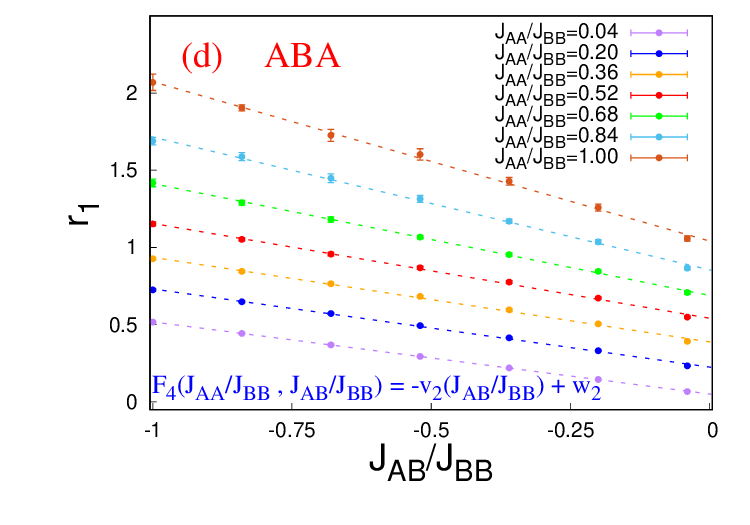}}
			
		\end{tabular}
		\caption{ (Colour Online) Behaviours of the fitting parameters of diluted ABA trilayered triangular stacking. Where the errorbars are not visible, they are smaller than the point markers.}
		\label{fig_15_aba_param}
	\end{center}
\end{figure*}

\begin{figure*}[!htb]
	\begin{center}
		\begin{tabular}{c}
			
			\resizebox{8.0cm}{!}{\includegraphics[angle=0]{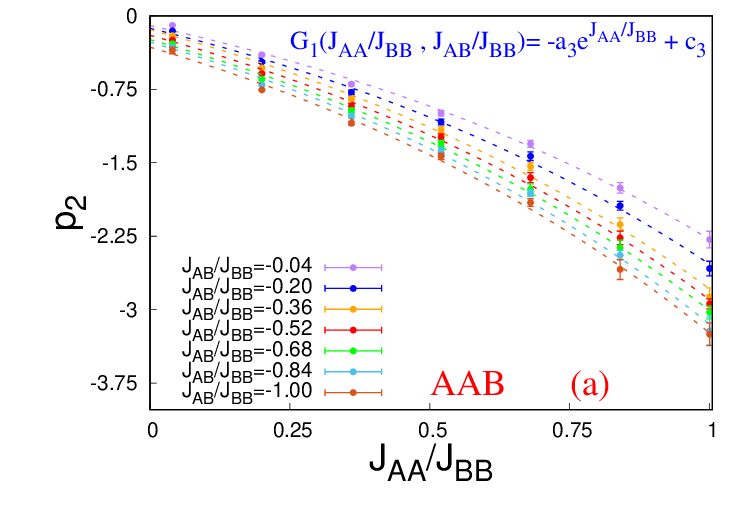}}
			\resizebox{8.0cm}{!}{\includegraphics[angle=0]{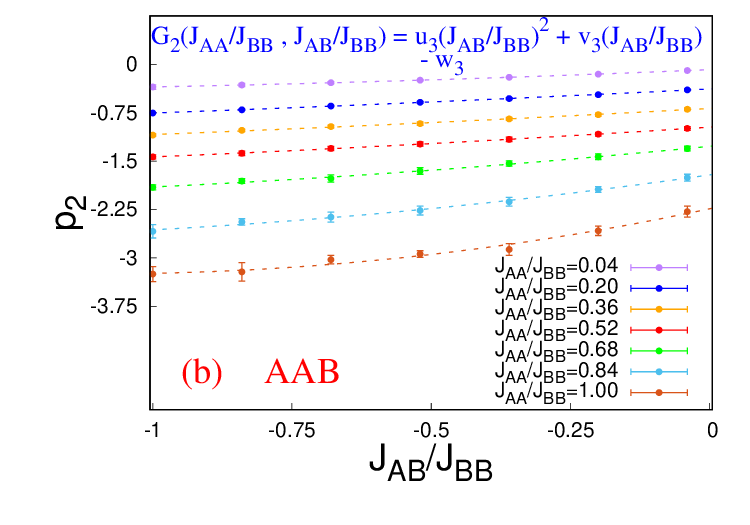}}\\
			
			\resizebox{8.0cm}{!}{\includegraphics[angle=0]{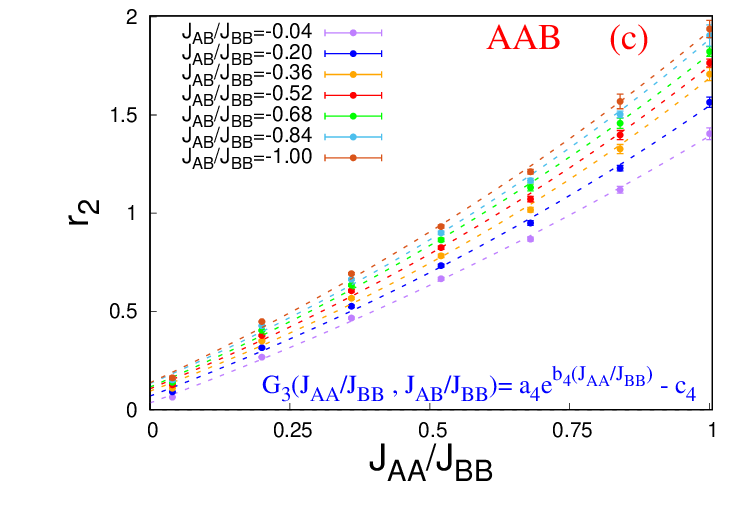}}
			\resizebox{8.0cm}{!}{\includegraphics[angle=0]{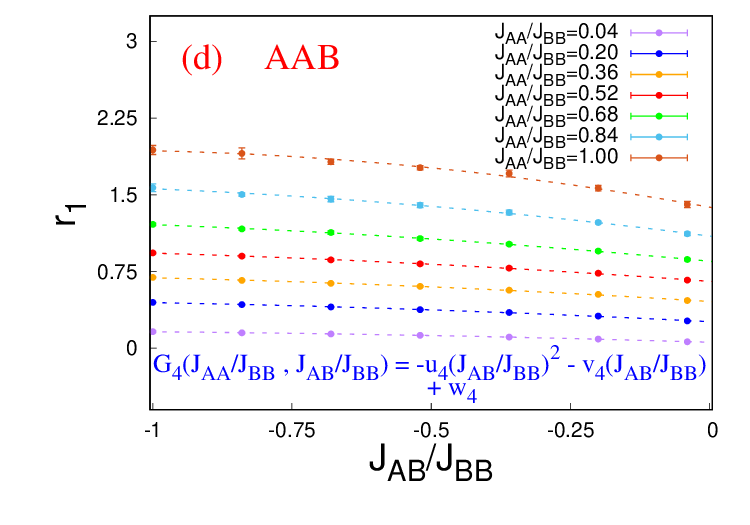}}
			
		\end{tabular}
		\caption{ (Colour Online) Behaviours of the fitting parameters of diluted AAB trilayered triangular stacking. Where the errorbars are not visible, they are smaller than the point markers.}
		\label{fig_16_aab_param}
	\end{center}
\end{figure*}

For the ABA type of trilayered stacking, in Figure \ref{fig_15_aba_param}, we see the parameters $p_{1}$ and $r_{1}$ can be modelled by simple functions: (a) exponential and (b) linear. For the AAB type of trilayered stacking, in Figure \ref{fig_16_aab_param}, the parameters $p_{2}$ and $r_{2}$ can be similarly modelled by: (a) exponential and (b) quadratic functions. All these choices are \textit{not unique} \cite{Press} but they provide us with a very simple yet complete way into the behaviours of the compensation temperatures for both configurations. The fitting formulae are all mentioned within the figures.
 
\section{On Compensation point}
\label{appendix_comp}
Now the pertinent question for similar types of studies is whether the compensation temperature is dependent on system size. We remind ourselves that the compensation point is that temperature below the critical point where the bulk magnetization of the system becomes zero i.e. $T_{comp}<T_{crit}$ and Compensation is not related to criticality \cite{Diaz1,Diaz2}. For the pristine counterpart of the system in this study it was shown in \cite{Chandra3} that for $L\geq 70$, the compensation temperature becomes independent of the size of the system. To recall how to detect the value of $T_{comp}$ in appropriate cases, we may look into Figure \ref{fig_2_mag_rho}. The intersection of $M_{tot}(T)$ with the x-axis (temperature axis/zero magnetisation line), below the critical point, is the Compensation point. While determining the point of intersection, we would approximate the part of the magnetization curve between the two neighbouring points on either side of the x-axis by a straight line \cite{Chandra2,Chandra3}.\\

\begin{figure*}[!htb]
	\begin{center}
		\begin{tabular}{c}
			
			\resizebox{8.0cm}{!}{\includegraphics[angle=0]{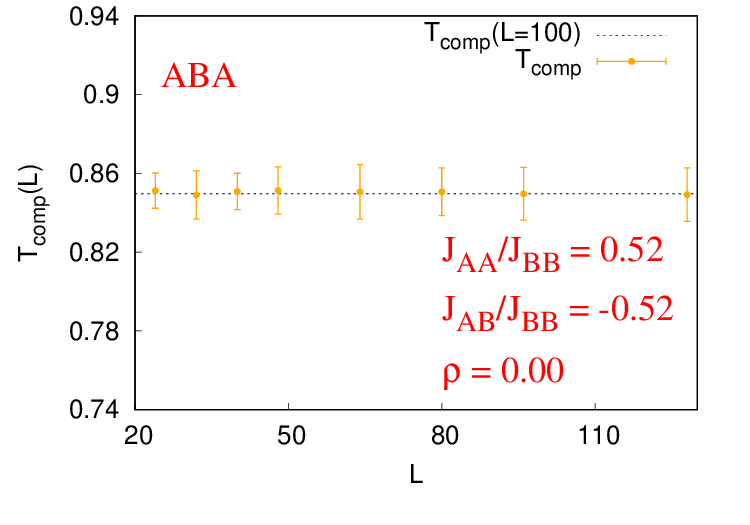}}
			\resizebox{8.0cm}{!}{\includegraphics[angle=0]{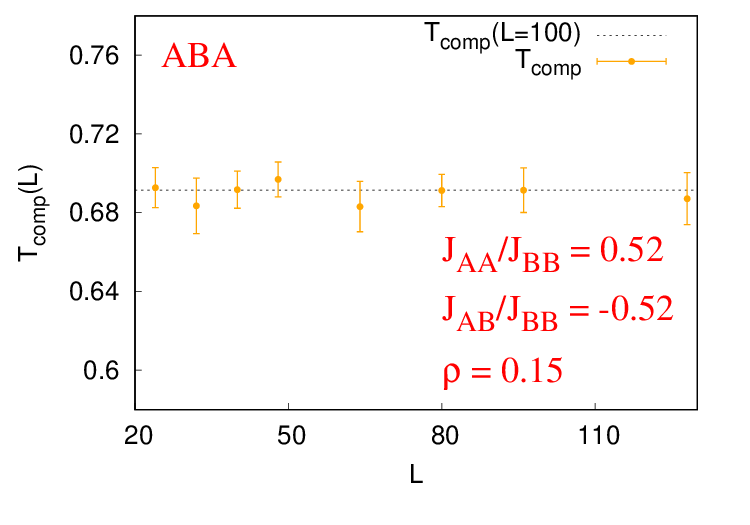}}\\
			
			\resizebox{8.0cm}{!}{\includegraphics[angle=0]{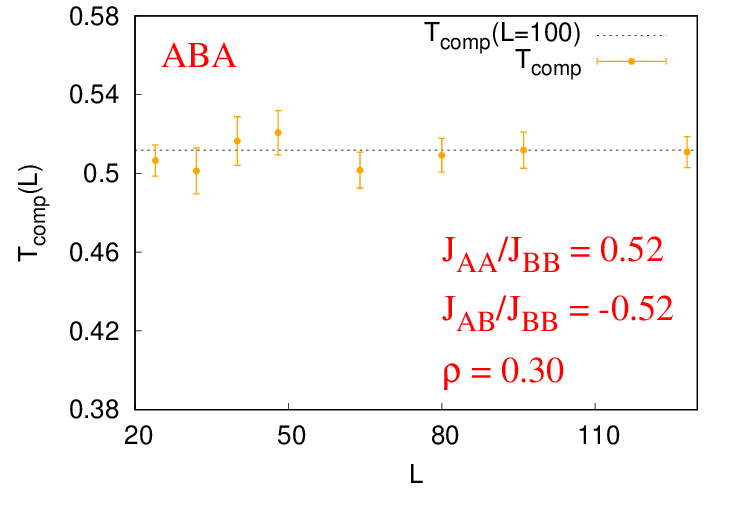}}
			\resizebox{8.0cm}{!}{\includegraphics[angle=0]{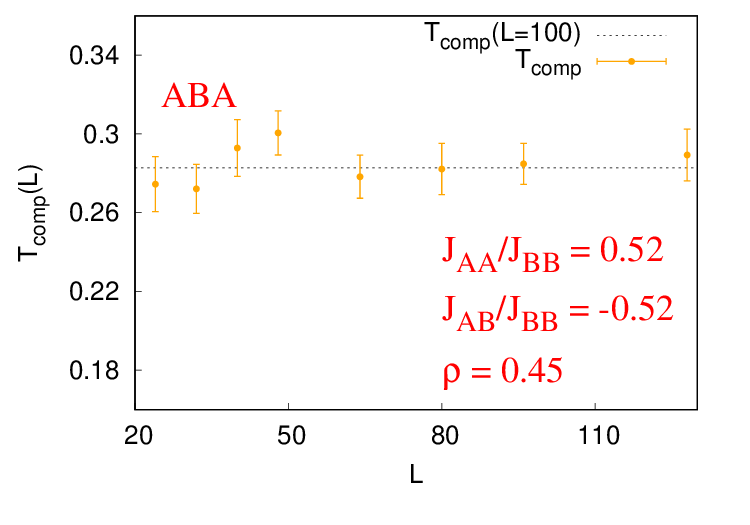}}
			
		\end{tabular}
		\caption{ (Colour Online) Compensation temperature versus linear system size of a diluted ABA trilayered triangular stacking. The reported value of compensation temperature ($L=100$) saturates about $L>64$.}
		\label{fig_17_aba_comp}
	\end{center}
\end{figure*}

\begin{figure*}[!htb]
	\begin{center}
		\begin{tabular}{c}
			
			\resizebox{8.0cm}{!}{\includegraphics[angle=0]{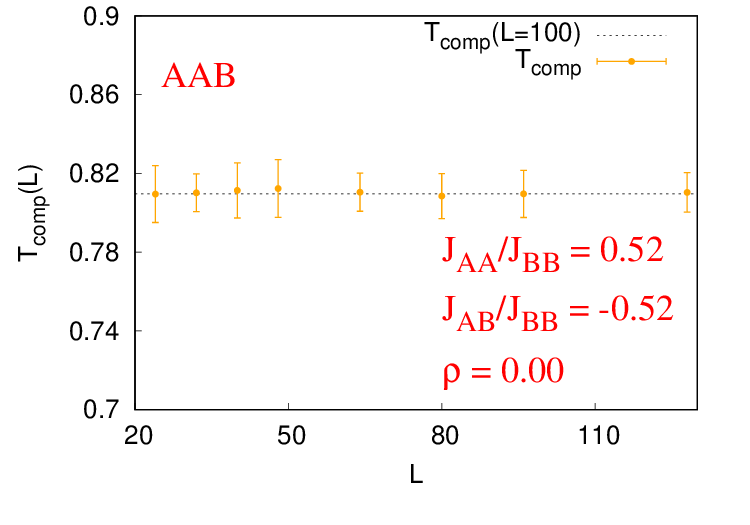}}
			\resizebox{8.0cm}{!}{\includegraphics[angle=0]{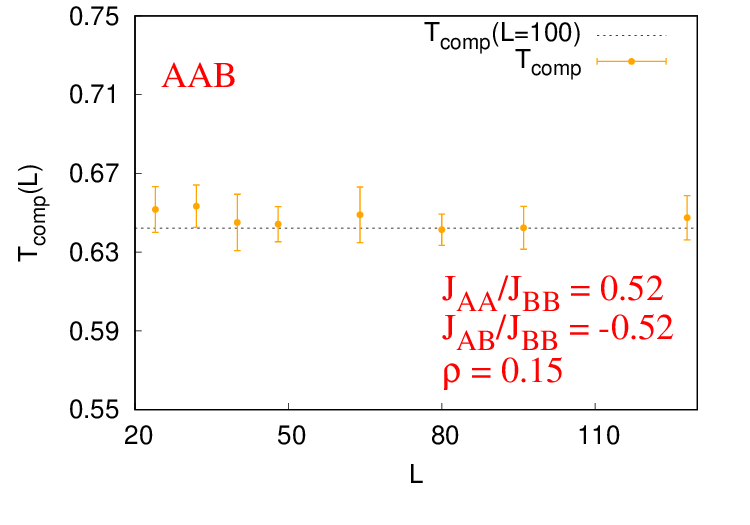}}\\
			
			\resizebox{8.0cm}{!}{\includegraphics[angle=0]{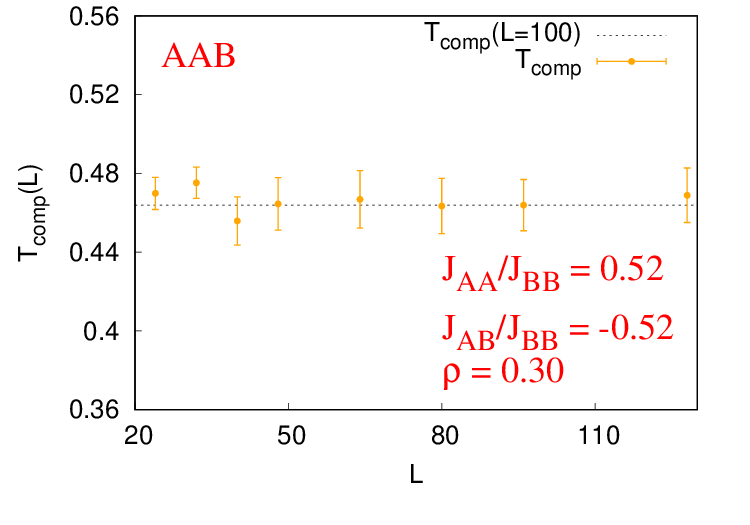}}
			\resizebox{8.0cm}{!}{\includegraphics[angle=0]{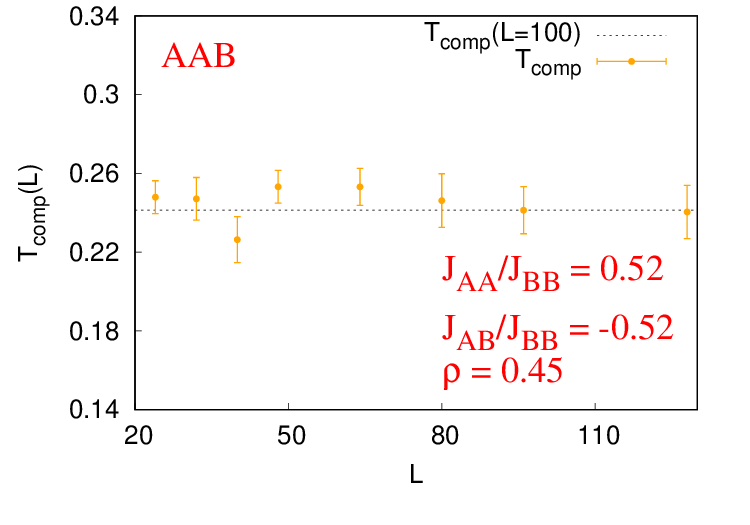}}
			
		\end{tabular}
		\caption{ (Colour Online) Compensation temperature versus linear system size of a diluted AAB trilayered triangular stacking. The reported value of compensation temperature ($L=100$) saturates about $L>64$.}
		\label{fig_18_aab_comp}
	\end{center}
\end{figure*}

In this study, we have varied the fraction of diluted sites from $0$ (pure sample) to $0.45$ . So one would naturally ask how the compensation point behaves as a function of linear system size, $L$, under the influence of growing nonmagnetic impurities. Because of the limited computational resources available, only a few random combinations of coupling strengths were tested in this regard. For both the ABA and AAB configurations, we present here the results for $J_{AA}/J_{BB}=0.52$ and $J_{AB}/J_{BB}=-0.52$ with four selective ratios of site-dilution $\rho=\{0.00,0.15,0.30,0.45\}$. This specific combination of the coupling ratios resides almost exactly at the middle of the investigated spectrum of the $J$'s and the equispaced values of $\rho$ enable us towards a systematic conclusion. From Figures \ref{fig_17_aba_comp} and \ref{fig_18_aab_comp}, we could see fluctuations in the values of compensation temperature for smaller system sizes, upto $L\lesssim64$, when we introduce and increase the dilution percentage in steps upto $45\%$. But after $L>64$, the fluctuations subside with the value of the compensation temperature becoming almost immune to the system size. For very few other randomly checked combinations of coupling strengths, the same feature is present. So, within the scope of available limited computational resources, we may only use the values of compensation temperatures for linear system sizes $L=100$ without compromising much on the accuracy for the entire $490$ different combinations of coupling strengths and dilution percentage. The discussion here may validate the compromise and choice of the system size used in this study.

\begin{enumerate}

\bibitem{Neel} 
N\'{e}el M. L., Ann. de Phys. \textbf{ 12}, 137 (1948).
	
\bibitem{Cullity}
Cullity B. D. and Graham C. D., Introduction to Magnetic Materials, 2nd edn. (Wiley, New York, 2008) .

\bibitem{Diaz1}
Diaz I. J. L. and Branco N. S., Phys. B: Condens. Matter \textbf{ 529}, 73 (2018) .

\bibitem{Diaz2}
Diaz I. J. L. and Branco N. S., Phys. A: Stat. Mech. Appl. \textbf{ 540}, 123014 (2020).

\bibitem{Chandra1}
Chandra S. and Acharyya M., AIP Conf. Proc. \textbf{ 2220}, 130037 (2020) .

\bibitem{Chandra2}
Chandra S., Eur. Phys. J. B \textbf{ 94}, 13 (2021) .

\bibitem{Chandra3}
Chandra S., J. Phys. Chem. Solids \textbf{ 156}, 110165 (2021) .

\bibitem{Chandra4}
Chandra S., Phys. Rev. E \textbf{ 104}, 064126 (2021) .

\bibitem{Chandra5}
Chandra S., arXiv:2201.03883 (2022) .

\bibitem{Camley}
Camley R. E. and Barna\'{s} J., Phys. Rev. Lett. \textbf{ 63}, 664 (1989).

\bibitem{Connell}
Connell G., Allen R. and Mansuripur M., J. Appl. Phys. \textbf{ 53}, 7759 (1982).

\bibitem{Phan}
Phan M.-H. and Yu S.-C., Journal of Magnetism and Magnetic Materials \textbf{ 308}, 325 (2007).

\bibitem{Grunberg}
Gr\"{u}nberg P., Schreiber R., Pang Y., Brodsky M. B. and Sowers H., Phys. Rev. Lett. \textbf{ 57}, 2442 (1986).

\bibitem{Stier}
Stier M. and Nolting W., Phys. Rev. B \textbf{ 84}, 094417 (2011).

\bibitem{Smits}
Smits C., Filip A., Swagten H., Koopmans B., De Jonge W., et al., Phys. Rev. B \textbf{ 69}, 224410 (2004).

\bibitem{Leiner}
Leiner J., Lee H., Yoo T., Lee S., Kirby B., et al., Phys. Rev. B \textbf{ 82}, 195205 (2010).

\bibitem{Chern}
Chern G., Horng L., Shieh W. K., and Wu T. C., Phys. Rev. B \textbf{ 63}, 094421 (2001).

\bibitem{Sankowski}
Sankowski P. and Kacman P., Phys. Rev. B \textbf{ 71}, 201303 (2005).

\bibitem{Chung}
Chung J.-H., Song Y.-S., Yoo T., Chung S. J., Lee S., et al., Journal of Applied Physics \textbf{ 110}, 013912 (2011).

\bibitem{Samburskaya}
Samburskaya T., Sipatov A. Y., Volobuev V., Dziawa P., Knoff W., et al., Acta Phys. Pol. A \textbf{ 124}, 133 (2013).

\bibitem{Pradhan}
Pradhan A., Maitra T., Mukherjee S., Mukherjee S., Nayak A., et al., Mater Lett. \textbf{ 210}, 77 (2018).

\bibitem{Maitra}
Maitra T., Pradhan A., Mukherjee S., Mukherjee S., Nayak A. and Bhunia S., Physica E \textbf{ 106}, 357 (2019).

\bibitem{Herman}
Herman M. A. and Sitter H., Molecular beam epitaxy: fundamentals and current status, Vol. 7 (Springer Science \& Business
Media, 2012).

\bibitem{Stringfellow}
Stringfellow G. B., Organometallic vapor-phase epitaxy: theory and practice (Academic Press, 1999).

\bibitem{SinghRK}
Singh R. K. and Narayan J., Phys. Rev. B \textbf{ 41}, 8843 (1990).


\bibitem{Leskela}
Leskel\"{a} M. and Ritala M., Angewandte Chemie International Edition \textbf{ 42}, 5548 (2003).

\bibitem{George}
George S. M., Chem. Rev \textbf{ 110}, 111 (2010).

\bibitem{Sandeman}
Sandeman K. G., Scr. Mater. \textbf{ 67}, 566 (2012).

\bibitem{Manosa}
Manosa L., Planes A. and Acet M., J. Mater. Chem. A \textbf{ 1}, 4925 (2013).

\bibitem{Amaral}
Amaral J. S., Silva N. J. O. and Amaral V. S., Appl. Phys. Lett. \textbf{ 91}, 172503 (2007).

\bibitem{Franco} 
Franco V., Conde A., Romero-Enrique J. M. and Bl\'{a}zquez J. S., J. Phys.: Condens. Matter \textbf{ 20}, 285207 (2008).

\bibitem{Dong} 
Dong Q. Y., Zhang H. W., Sun J. R., Shen B. G. and Franco V., J. Appl. Phys. \textbf{ 103}, 116101 (2008).

\bibitem{Oliveira1} 
de Oliveira N. and von Ranke P., Phys. Rep. \textbf{ 489}, 89 (2010).

\bibitem{Amaral2} 
Amaral J. S. and Amaral V. S. in : \textit{Thermodynamics: Systems
in Equilibrium and Non-Equilibrium}, ed. Moreno-Piraj\'{a}n J. C., chapter 8, pp. 173-198 (2011).

\bibitem{Franco2}
Franco V., Blazquez J. S., Ingale B. and Conde A., Annu. Rev. Mater. Res. \textbf{ 42}, 305 (2012).

\bibitem{Pelka}
Pe\l{}ka R., et al., Acta Phys. Pol. A \textbf{ 124}, 977 (2013).

\bibitem{Canova}
\v{C}anov\'{a} L., Stre\v{c}ka J. and Ja\v{s}\v{c}ur M. 2006 J. Phys.: Condens. Matter \textbf{ 18}, 4967 (2006).

\bibitem{Pereira}
Pereira M. S. S., de Moura F. A. B. F. and Lyra M. L., Phys. Rev. B \textbf{ 79}, 054427 (2009).

\bibitem{Ohanyan} 
Ohanyan V. and Honecker A., Phys. Rev. B \textbf{ 86}, 054412 (2012).


\bibitem{Galisova} 
G\'{a}lisov\'{a} L., Condens. Matter Phys. \textbf{ 17}, 13001 (2014).


\bibitem{Ribeiro} 
Ribeiro G. A. P., J. Stat. Mech. \textbf{ 2010}, P12016 (2010).

\bibitem{Trippe}
Trippe C., Honecker A., Kl\"{u}mper A. and Ohanyan V., Phys. Rev. B \textbf{ 81}, 054402 (2010).

\bibitem{Zhitomirsky}
Zhitomirsky M. E. and Honecker A. 2004 J. Stat. Mech. \textbf{ 2004}, P07012 (2004).

\bibitem{Topilko}
Topilko M., Krokhmalskii T., Derzhko O. and Ohanyan V., Eur. Phys. J. B \textbf{ 85}, 1 (2012).

\bibitem{Chang}
Chang C., Wang W., Lv D., Liu Z. and Tian M.,
Eur. Phys. J. Plus \textbf{ 136}, 290 (2021).

\bibitem{Guru}
Guru E. and Saha S., Phase Transit., (2022);\\ \textit{DOI: 10.1080/01411594.2022.2095274} .

\bibitem{Oliveira2}
Oliveira S., Morais R. H. M., Santos J. P. and S\'{a} Barreto F. C., Condens Matter Phys., \textbf{ 25(1)}, 13702 (2022).







\bibitem{Aydiner}
Aydiner E., Y\"{u}ksel Y., Kis-Cam E., and Polat H., J. Magn. Magn. Mater. \textbf{ 321}, 3193 (2009).

\bibitem{Yuksel}
Y\"{u}ksel Y., Akinci \"{U}., and Polat H., Phys. Status Solidi B \textbf{ 250(1)}, 196 (2013). 

\bibitem{Diaz3}
Diaz I. J. L. and Branco N. S., Phys. A \textbf{ 468}, 158 (2017).

\bibitem{Vatansever}
Vatansever E. and Y\"{u}ksel Y., J. Magn. Magn. Mater. \textbf{ 441}, 548 (2017).

\bibitem{Diaz4}
Diaz I. J. L. and Branco N. S., Physica A \textbf{ 490}, 904 (2018). 

\bibitem{Fadil}
Fadil Z., Qajjour M., Mhirech A., Kabouchi B., Bahmad L., et al., Physica B \textbf{ 564}, 104 (2019).

\bibitem{Qajjour}
Qajjour M., Maaouni N., Fadil Z., Mhirech A., Kabouchi B., et al., Chin. J. Phys. \textbf{ 63}, 36 (2020).

\bibitem{Buluta}
Buluta I., and Nori F., Science \textbf{ 326}, 108 (2009).

\bibitem{Lewenstein}
Lewenstein M., et al., Adv. Phys. \textbf{ 56}, 243 (2007).

\bibitem{Struck}
Struck J., et al., Science \textbf{ 333}, 996 (2011).

\bibitem{Britton}
Britton J. W., Sawyer B. C., Keith A. C., Wang C.-C.J., Freericks J. K., et al., Nature \textbf{ 484}, 489 (2012).

\bibitem{Biercuk}
Biercuk M. J., et al., Quant. Inf. Comput. \textbf{ 9}, 920 (2009).

\bibitem{Ising}
Ising E., Z. Phys. \textbf{ 31}, 253 (1925).

\bibitem{Landau-Binder}
Landau D. P. and Binder K., A Guide to Monte Carlo Simulations in Statistical Physics (Cambridge University Press, New York, 2000).

\bibitem{Binder-Heermann}
Binder K. and Heermann D. W., Monte Carlo Simulation in Statistical Physics (Springer, New York, 1997).

\bibitem{Ferrenberg}
Ferrenberg A. M. and Landau D. P., Phys. Rev. B \textbf{44 (10)}, 5081 (1991).

\bibitem{Newman}
Newman M. E. J. and Barkema G. T., Monte Carlo Methods in Statistical Physics (Oxford University Press, New York, 2001).

\bibitem{Strecka}
Stre\v{c}ka J., Physica A \textbf{ 360}, 379 (2006).

\bibitem{Chikazumi}
Chikazumi S., Physics of Ferromagnetism (Oxford University Press, Oxford, 1997).

\bibitem{Miyazaki}
Miyazaki T. and Jin H., The Physics of Ferromagnetism (Springer Series in Materials Science 158, 2012)

\bibitem{Scarborough}
Scarborough J.B., Numerical Mathematical Analysis, (Oxford \& Ibh, London, 2005).

\bibitem{Krauth}
Krauth W., Statistical Mechanics: Algorithms and Computations (Oxford University Press, UK, 2006)

\bibitem{Press}
Press W. H., Teukolsky S. A., Vetterling W. T. and Flannery B. P., Numerical Recipes: The Art of Scientific Computing (Cambridge University Press, Cambridge, UK, 2007)

\end{enumerate}
\end{document}